\documentclass[prd,onecolumn,reprint,superscriptaddress]{revtex4-2}
\usepackage{feynmp,amsmath,amssymb,graphicx,subfigure,hyperref,bm,mathrsfs,slashed}
\usepackage{color}

\begin{document}

\title{Interband optical conductivities in two-dimensional tilted Dirac bands revisited within the tight-binding model}

\author{Chao-Yang Tan}
\thanks{These authors have contributed equally to this work.}
\affiliation{Department of Physics and
Center for Computational Sciences, Sichuan Normal University, Chengdu,
Sichuan 610066, China}
\affiliation{Department of Physics and Beijing Key Laboratory of Opto-electronic Functional Materials and Micro-nano Devices, Renmin University of China, Beijing 100872, China}

\author{Jian-Tong Hou}
\thanks{These authors have contributed equally to this work.}
\affiliation{Department of Mathematics,
	The Hong Kong University of Science and Technology, Kowloon, Hong Kong, China}
\affiliation{Department of Physics, Institute of Solid State Physics and Center for Computational Sciences, Sichuan Normal University, Chengdu, Sichuan 610066, China}

\author{Xin Chen}
\thanks{These authors have contributed equally to this work.}
\affiliation{Department of Physics and
	Center for Computational Sciences, Sichuan Normal University, Chengdu,
	Sichuan 610066, China}

\author{Ling-Zhi Bai}
\affiliation{Department of Physics and
	Center for Computational Sciences, Sichuan Normal University, Chengdu,
	Sichuan 610066, China}

\author{Jie Lu}
\affiliation{College of Physics Science and Technology, Yangzhou University, Yangzhou 225002, China}

\author{Yong-Hong Zhao}
\affiliation{Department of Physics and
	Center for Computational Sciences, Sichuan Normal University, Chengdu,
	Sichuan 610066, China}

\author{Chang-Xu Yan}
\thanks{Corresponding author: cxyan@mail.bnu.edu.cn}
\affiliation{Department of Physics and
	Center for Computational Sciences, Sichuan Normal University, Chengdu,
	Sichuan 610066, China}
\affiliation{School of Physics and Astronomy, Beijing Normal University, Beijing 100875, China}

\author{Hao-Ran Chang}
\thanks{Corresponding author:hrchang@mail.ustc.edu.cn}
\affiliation{Department of Physics and
	Center for Computational Sciences, Sichuan Normal University, Chengdu,
	Sichuan 610066, China}

\author{Hong Guo}
\affiliation{Department of Physics, McGill University, Montreal, Quebec H3A 2T8, Canada}
\affiliation{Department of Physics and
	Center for Computational Sciences, Sichuan Normal University, Chengdu,
	Sichuan 610066, China}

\date{\today}

\begin{abstract}
Within the framework of linear response theory, we theoretically investigated the interband longitudinal optical conductivities (LOCs) in two-dimensional (2D) tilted Dirac bands using a tight-binding (TB) model, incorporating the effects of band tilting and Dirac-point shifting. We identified three characteristic critical frequencies in the interband LOCs of the TB model: the partner frequencies, the sharp- peak frequency, and the cutoff frequency. In contrast to conventional critical frequencies, these three types are consistently absent in the corresponding linearized $k\cdot p$ model. Notably, the sharp-peak frequency and cutoff frequency remain robust against variations in band tilting and Dirac-point shifting. By employing analytical expressions derived via the Lagrange multiplier method, we elucidate the origins of the conventional critical frequencies and their partner counterparts. In contrast, the sharp-peak frequency and cutoff frequency are associated with interband optical transitions at high-symmetry points of the energy bands, arising from the Pauli exclusion principle and the finite boundaries of the Brillouin zone. Our theoretical predictions are intended to guide future experimental studies on tilt-dependent optical phenomena in 2D tilted Dirac systems.
\end{abstract}

\maketitle

\section{Introduction\label{Sec:intro}}

Two-dimensional (2D) Dirac materials, characterized by linearly dispersing Dirac bands around Dirac points in momentum space, have attracted great and sustained attention since the exfoliation of graphene \cite{Science2004,RMP2009}. Their Dirac bands can be tilted along a specific wave-vector direction, introducing intrinsic anisotropy into the energy dispersion. Such 2D tilted Dirac bands have been studied theoretically and experimentally in a series of materials, including $\alpha$-(BEDT-TTF)$_2$I$_3$ \cite{JPSJ2006}, graphene under uniaxial strain \cite{ChoiPRB2010}, 8-$\textit{Pmmn}$ borophene \cite{Zhou8Pmmn2014PRL,Science8Pmmn2015,PRB8PmmnRapid2016,PRB8Pmmn2016}, transition metal dichalcogenides \cite{PRLMoS2010,PRLMoS2012,Science2014}, partially hydrogenated graphene \cite{TingPRB2016}, $\alpha$-SnS$_2$ \cite{NPGMa2016}, graphdiyne \cite{PRLYang2019}, TaCoTe$_2$ \cite{PRBYang2019}, and TaIrTe$_4$ \cite{PRBLu2020}. Compared to their untilted counterparts, these 2D tilted Dirac materials exhibit a wide range of qualitatively distinct physical behaviors, such as plasmons \cite{PRBIurov2017,PRBAgarwal2017,PRBJafari2018,CPBLiu2022,PRBMojarro2022,JPSJNishine2011,JPSJNishine2010}, optical conductivities \cite{JPSJNishine2010,PRBVerma2017,PRBIurov2018,PRBHerrera2019,PRBRostamzadeh2019,PRBTan2021,PRBTan2022,PRBHou2023,PRBMojarro2021,PRBWild2022,PRBYao2021}, Weiss oscillation \cite{PRBIslam2017}, Klein tunneling \cite{PRBSHZhang2018,NanomaterialsKongPRB2018,PRBNguyen2018}, Kondo effects \cite{PRBSun2018}, RKKY interactions \cite{PRBPaul2019,JMMMZhang2019}, planar Hall effect \cite{PRBWang2020R,PRBRostami2020}, valley Hall effect \cite{PRLZhang2023}, thermoelectric effects \cite{PRBKapri2020}, thermal currents \cite{PRBSengupta2020}, valley filtering \cite{PRBZheng2021}, gravitomagnetic effects \cite{PRBFarajollahpour2020}, Andreev reflection \cite{PRBFaraei2020}, Coulomb bound states \cite{PRBFu2021}, guided modes \cite{SciRepNg2022}, and valley-dependent time evolution of coherent electron states \cite{PRBYonatan2022}.

As an essential experimental probe, optical conductivity is highly sensitive to energy bands and can be used to extract key information on the band structures and optical properties of materials \cite{RMP2009}. Due to the intrinsic anisotropy of 2D tilted Dirac bands, their optical conductivities exhibit strongly anisotropic behavior. Consequently, the optical conductivities in 2D tilted Dirac bands \cite{JPSJNishine2010,PRBVerma2017,PRBIurov2018,PRBHerrera2019,PRBRostamzadeh2019,PRBTan2021,PRBTan2022,PRBHou2023,PRBMojarro2021,PRBWild2022,PRBYao2021} differ qualitatively from those in untilted 2D Dirac bands \cite{PRLCarbotte2006,PRBGusynin2007,PRLMikhailov2007,PRLKuzmenko2008,PRLMak2008,PRBStauber2008,PRBStille2012,PRBCarbotte2012,PRBCarbotte2013,PRBXiao2013,RPWu2018,PLAWu2019,EPJBWu2019}. However, most of these studies were conducted within the framework of the linearized $k \cdot p$ Hamiltonian. To assess whether the linearized $k \cdot p$ Hamiltonian adequately captures the optical properties of these 2D Dirac systems, we revisit the interband longitudinal optical conductivities (LOCs) using the tight-binding (TB) Hamiltonian for 2D tilted Dirac bands and perform a comprehensive comparison with results obtained from the linearized $k \cdot p$ Hamiltonian.

Within the linear response theory, we theoretically investigate the LOCs in 2D tilted Dirac bands using the TB model, incorporating the effects of band tilting and Dirac-point shifting. We identify three characteristic critical frequencies in the interband LOCs of the TB model: the partner frequencies, the sharp-peak frequency, and the cutoff frequency. In contrast to conventional critical frequencies, these three types are consistently absent in the corresponding linearized $k \cdot p$ model. We explain the origins of these characteristic critical frequencies using both analytical expressions derived either via the Lagrange multiplier method or by analyzing interband optical transitions at high-symmetry points of the energy bands. Our theoretical predictions can guide future experimental studies of tilt- dependent phenomena in the optical measurement of 2D tilted Dirac materials.

The rest of this paper is organized as follows. In Section \ref{Sec:Model}, we briefly outline the model Hamiltonian and the theoretical formalism used to calculate the interband LOCs. Section \ref{Sec:results and analysis} presents our numerical results and corresponding analytical expressions. In Section \ref{Sec:comparisons}, we compare the interband LOCs obtained from the TB model with those derived from the linearized $k \cdot p$ model. Finally, our conclusions and discussions are provided in Section \ref{Sec:conclusions and discussions}.

\section{Theoretical formalism in the tight-binding Hamiltonian \label{Sec:Model}}

\begin{figure}[htbp]
	\centering
	\includegraphics[width=8cm]{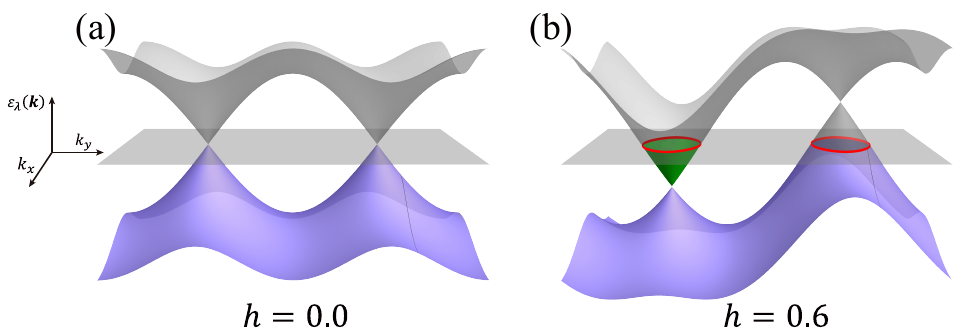}
	\caption{Schematic diagrams of energy band for (a) $h=0$ and (b) $h=0.6$.}
	\label{fig1}
\end{figure}

We begin with a TB Hamiltonian for 2D tilted Dirac fermion
\begin{align}
	\mathcal{H}(\boldsymbol{k})
	=\mathcal{H}(k_x,k_y)
	&=\left[t\cos(a k_y)+h\sin(a k_y)\right]\varepsilon_{0}\tau_0
	\nonumber\\&
	\hspace{+0.25cm}
	+\sin(a k_x)\varepsilon_{0}\tau_1 +\cos(a k_y)\varepsilon_{0}\tau_2,
	\label{Eq1}
\end{align}
where $a$ denotes the lattice constant, $\boldsymbol{k}=(k_x, k_y)$ stands for the wave vector, and $\tau_0$ and $\tau_i$ are the $2\times2$ unit matrix and Pauli matrices, respectively. The parameter $t$ quantifies the band tilting along the $k_y$-direction, which breaks the isotropic nature of the Dirac cone and leads to anisotropic energy dispersion. The parameter $h$ breaks time-reversal symmetry and also contributes to the breaking of spatial inversion symmetry in conjunction with the $t$ term. The parameter $\varepsilon_{0}$ represents the energy scale of validity for Dirac linear dispersion, which falls typically within $0.2\sim 2.0~\mathrm{eV}$ for many real 2D Dirac materials \cite{footnote1}. A straightforward diagonalization yields the eigenvalues of the TB Hamiltonian in Eq.(\ref{Eq1}) as
\begin{align}
	\varepsilon_{\lambda}(k_x,k_y)
	&=t\cos(a k_y)\varepsilon_{0}+h\sin(a k_y)\varepsilon_{0}
	\nonumber\\&
	\hspace{+0.25cm}
	+\lambda\mathcal{Z}(k_x,k_y)\varepsilon_{0}
	,\label{eigenvalue}
\end{align}
where
\begin{align}
	\mathcal{Z}(k_x,k_y)=\sqrt{\sin^2(a k_x)+\cos^2(a k_y)},
\end{align}
and $\lambda=\pm$ denotes the conduction and valence bands, respectively.
In the vicinity of the two Dirac points $(0,-\kappa\frac{\pi}{2a})$, the TB Hamiltonian in Eq.(\ref{Eq1}) describes a pair of oppositely tilted Dirac fermions, where $\kappa=+$ and $\kappa=-$ label the left valley around Dirac point $(0,-\frac{\pi}{2a})$ and the right valley around Dirac point $(0,+\frac{\pi}{2a})$ in Fig.\ref{fig1}, respectively.

In the untilted phase ($t=0$), the eigenvalues are explicitly shown in Fig.~\ref{fig1}, in which panel (a) and panel (b) correspond to $h=0$ and $h\neq0$, respectively. Obviously, the Dirac points at two separated valleys are on the different zero-energy when $h\neq0$; in contrast to the same zero-energy when $h=0$, in which $h$ measures the energy-shifting of Dirac points with respect to the \emph{initial} position of $h=0$. The two Dirac points in Fig.~\ref{fig1} can also be further tilted oppositely. The interplay between $t$ and $h$ thus governs the symmetry-breaking characteristics of the system, which in turn significantly influence its optical and electronic responses. In the following, we restrict our analysis to untilted phase ($t=0$) and under-tilted phase ($0<t<1$), respectively.

Introducing the electromagnetic field via minimal coupling and expanding the TB Hamiltonian to the leading order of $e$ (see the Supplemental Materials \cite{SupplementalMaterials}), we have 
\begin{align}
\mathcal{H}(\boldsymbol{k}+e\boldsymbol{A})
&=\mathcal{H}(\boldsymbol{k})
\nonumber\\&
+eaA_x\cos(ak_x)\varepsilon_{0}\tau_1-eaA_y\sin(ak_y)\varepsilon_{0}\tau_2
\nonumber\\&
-eaA_y\left[t\sin(ak_y)-h\cos(ak_y)\right]\varepsilon_{0}\tau_0.\nonumber
\end{align}
Consequently, the current operators $\hat{J}_j$ with $j=x,y$ are obtained as
\begin{align}
\hat{J}_x&=\frac{\partial \mathcal{H}(\boldsymbol{k}+e\boldsymbol{A})}{\partial A_x}
=ea\cos(ak_x)\varepsilon_{0}\tau_1, \label{current_x}\\
\hat{J}_y&=\frac{\partial \mathcal{H}(\boldsymbol{k}+e\boldsymbol{A})}{\partial A_y}
\nonumber\\&
=-ea\sin(ak_y)\varepsilon_{0}(t\tau_0+\tau_2)+ea\cos(ak_y)\varepsilon_{0}h\tau_0.
\label{current_y}
\end{align}

Within linear response theory, the LOCs $\sigma_{jj}(\omega,\mu,t,h)$ at finite photon frequency $\omega$ can be given in terms of the current-current correlation function $\Pi_{jj}(\omega,\mu,t,h)$ as
\begin{align}
\sigma_{jj}(\omega,\mu,t,h)&=\frac{\mathrm{i}}{\omega}\Pi_{jj}(\omega,\mu,t,h),
\end{align}
where $\mu$ measures the chemical potential with respect to the Dirac point. The current-current correlation function is defined by
\begin{align}
	&\Pi_{ij}(\omega,\mu,t,h)
	=\lim_{\boldsymbol{q}\to 0}\frac{1}{\beta}\sum_{i\Omega_m}
	\int_{-\infty}^{+\infty}\int_{-\infty}^{+\infty}\frac{\mathrm{d}^2\boldsymbol{k}}{(2\pi)^2}
	\nonumber\\&\hspace{0.5cm}
	\mathrm{Tr}\left[\hat{J}_i G(\boldsymbol{k},i\Omega_{m}) \hat{J}_j G(\boldsymbol{k}+\boldsymbol{q},i\Omega_{m}+\omega+i\eta)\right],
	\label{Pi}
\end{align}
where $\eta$ stands for a positive infinitesimal, $\beta$ denotes $1/(k_B T)$ with $k_B$ being the Boltzmann constant and $T$ the temperature, and the Matsubara Green's function takes the form (see the Supplemental Materials \cite{SupplementalMaterials})
\begin{align}
	G(\boldsymbol{k},i\Omega_{m})=&\left[(i\Omega_m+\mu)\tau_0-\mathcal{H}(\boldsymbol{k})\right]^{-1}
	\nonumber\\&
	=\frac{1}{2}\sum_{\lambda=\pm} \frac{\mathcal{P}_\lambda(\boldsymbol{k})}{i\Omega_m+\mu-\varepsilon_\lambda(k_x,k_y)},
\end{align}
with $i\Omega_m$ the Matsubara frequency and
\begin{align}
	\mathcal{P}_\lambda(\boldsymbol{k})&=\tau_0 +\lambda\frac{\tau_1\sin(ak_x)+\tau_2\cos(ak_y)}{\mathcal{Z}(k_x,k_y)}.
\end{align}

Hereafter, we focus on the interband optical transition between the valence band and the conduction band. For convenience, we restrict to the case where $\mu\ge0$. After some tedious but straightforward algebra (see the Supplemental Materials \cite{SupplementalMaterials}), the real part of the interband LOCs takes the form
\begin{align}
\mathrm{Re}~\sigma_{jj}^{\mathrm{IB}}(\omega,\mu,t,h)
&=e^2 \pi \int_{-\frac{\pi}{2a}}^{+\frac{\pi}{2a}} \frac{\mathrm{d} k_x}{2\pi} \int_{-\frac{\pi}{a}}^{+\frac{\pi}{a}}
 \frac{\mathrm{d} k_y}{2\pi} 
 \nonumber\\&\hspace{-0.5cm}\times
 \mathcal{F}_{jj}^{-,+}(k_x,k_y)
\delta\left[\omega-2\mathcal{Z}(k_x,k_y)\varepsilon_0\right]
\nonumber\\&\hspace{-0.5cm}\times
\frac{f\left[\varepsilon_{-}(k_x,k_y)\right]-f\left[\varepsilon_{+}(k_x,k_y)\right]} {\omega},
\label{integrationReg1}
\end{align}
where 
\begin{align}
\mathcal{F}_{xx}^{-,+}(k_x,k_y)
&=\frac{(a\varepsilon_0)^2\cos^2(a k_x) \cos^2(ak_y)}{\left[\mathcal{Z}(k_x,k_y)\right]^2}, \\
\mathcal{F}_{yy}^{-,+}(k_x,k_y)
&=\frac{(a\varepsilon_0)^2\sin^2(ak_x)\sin^2(ak_y)}{\left[\mathcal{Z}(k_x,k_y)\right]^2}.
\end{align}
In addition,  $f(x)=1/\left\{1+\exp\left[(x-\mu)/(k_B T)\right]\right\}$ denotes the Fermi distribution function, $\delta(x)$ the Dirac $\delta$-function, and $\sigma_0=\frac{e^2}{4\hbar}$ (we restore $\hbar$ for explicitness temporarily for explicitness). \emph{Hereafter, we use $\sigma_{jj}^{\mathrm{IB}}(\omega,\mu,t,h)$ instead of $\mathrm{Re}~\sigma_{jj}^{\mathrm{IB}}(\omega,\mu,t,h)$ to simplify the notation}.

\section{Results and analysis \label{Sec:results and analysis}}

The interband optical transitions are strongly related to the values of band tilting $t$, doping $\mu$, and shifting $h$ in Dirac materials. The angular dependence for interband LOCs are given in Subsection \ref{SubSecA}. The interband LOCs in the TB model and the corresponding analysis of characteristic critical frequencies for the untilted case ($t=0$) and under-tilted case ($0<t<1$) are mainly presented in the Subsections \ref{SubSecB} and \ref{SubSecC}, respectively. We further discuss the characteristic critical frequencies of the interband LOCs in Subsection \ref{SubSecD}. Throughout the numerical calculation of interband LOCs in the TB model, the temperature is set to be $T=1~\mathrm{K}$.

\subsection{Angular dependence for interband LOCs \label{SubSecA}}

To better illustrate the angular dependence of the Fermi wave vectors, conventional critical frequencies ($\omega_{j}$ or $\omega_{j}^{\pm}$), and their partner frequencies ($\omega_{j}^{\prime}$) for a given chemical potential $\mu$, we expand the TB Hamiltonian in Eq.(\ref{Eq1}) with respect to two Dirac points labeled by the valley index $\kappa=\pm$ as
\begin{align}
	\tilde{\mathcal{H}}_{\kappa}\left(\tilde{k}_x,\tilde{k}_y\right)&=
	\mathcal{H}\left(\tilde{k}_x,\tilde{k}_y-\kappa\frac{\pi}{2a}\right)
	\nonumber\\&
	=\left[\kappa t\sin(a \tilde{k}_y)-\kappa h\cos(a \tilde{k}_y)\right]\varepsilon_{0}\tau_0
	\nonumber\\&
	\hspace{+0.25cm}
	+\sin(a \tilde{k}_x)\varepsilon_{0}\tau_1 +\kappa \sin(a \tilde{k}_y)\varepsilon_{0}\tau_2,
	\label{TB2}
\end{align}
whose corresponding eigenenergy 
\begin{align}
	\tilde{\varepsilon}_{\lambda}^{\kappa}(\tilde{k}_x,\tilde{k}_y)
	&=\kappa t\sin(a\tilde{k}_y)\varepsilon_{0}-\kappa h\cos(a\tilde{k}_y)\varepsilon_{0}
	\nonumber\\&
	+\lambda\tilde{\mathcal{Z}}\left(\tilde{k}_x,\tilde{k}_y\right)\varepsilon_{0},
	\label{left}
\end{align}
with
\begin{align}
	\tilde{\mathcal{Z}}\left(\tilde{k}_x,\tilde{k}_y\right)=\sqrt{\sin^2(a \tilde{k}_x)+\sin^2(a\tilde{k}_y)},
\end{align}
where 
\begin{align}
	\tilde{k}_{x}\equiv \tilde{k} \cos\tilde{\theta}_{k} &\in \left[-\frac{\pi}{2a},+\frac{\pi}{2a}\right],\\
	\tilde{k}_{y}\equiv \tilde{k} \sin\tilde{\theta}_{k}&\in \left[-\frac{\pi}{2a},+\frac{\pi}{2a}\right],
\end{align}
are two components of the wave vector $\tilde{\boldsymbol{k}}=\left(\tilde{k}_{x},\tilde{k}_{y}\right)$. 
In terms of the variables $\tilde{\theta}_{k}$ and $\tilde{k}$, the eigenenergy can be rewritten as
\begin{align}
	&\hspace{-0.2cm}\tilde{\varepsilon}_{\lambda}^{\kappa}\left(\tilde{k}\cos\tilde{\theta}_{k},\tilde{k}\sin\tilde{\theta}_{k}\right)
	=\kappa t\sin\left(a\tilde{k}\sin\tilde{\theta}_{k}\right)\varepsilon_{0}
	\nonumber\\&\hspace{-0.2cm}
	-\kappa h\cos\left(a\tilde{k}\sin\tilde{\theta}_{k}\right)\varepsilon_{0}
	+\lambda\tilde{\mathcal{Z}}\left(\tilde{k}\cos\tilde{\theta}_{k},\tilde{k}\sin\tilde{\theta}_{k}\right)\varepsilon_{0}.
	\label{left}
\end{align}

For an arbitrary chemical potential $\mu$, the corresponding anisotropic Fermi wave vectors $\tilde{k}_{F}^{\kappa;\lambda}(\tilde{\theta}_{k})$—defined to be positive for all $\tilde{\theta}_{k}$—satisfy the equation
\begin{align}
	\tilde{\varepsilon}_{\lambda}^{\kappa}\left[\tilde{k}_{F}^{\kappa;\lambda}(\tilde{\theta}_{k})\cos\tilde{\theta}_{k},\tilde{k}_{F}^{\kappa;\lambda}(\tilde{\theta}_{k})\sin\tilde{\theta}_{k}\right]
	= \mu.
	\label{EqConstrain}
\end{align}
Interband optical transitions from the valence band to the conduction band can occur only when the photon energy satisfies the inequality
\begin{align}
	\omega &= 2\tilde{\mathcal{Z}}\left(\tilde{k}\cos\tilde{\theta}_{k},\tilde{k}\sin\tilde{\theta}_{k}\right)\varepsilon_{0}
	\nonumber\\&
	\geq \xi\left\{\mu-\tilde{\varepsilon}_{-\xi}^{\kappa}\left[\tilde{k}_{F}^{\kappa;\xi}(\tilde{\theta}_{k})\cos\tilde{\theta}_{k},\tilde{k}_{F}^{\kappa;\xi}(\tilde{\theta}_{k})\sin\tilde{\theta}_{k}\right]\right\}
	\nonumber\\&
	= 2\tilde{\mathcal{Z}}\left[\tilde{k}_{F}^{\kappa;\xi}(\tilde{\theta}_{k})\cos\tilde{\theta}_{k},\tilde{k}_{F}^{\kappa;\xi}(\tilde{\theta}_{k})\sin\tilde{\theta}_{k}\right]\varepsilon_{0},
	\label{EqTransition}
\end{align}
where $\xi=\mathrm{sgn}\left(\mu_{\kappa}\right)$ with $\mu_{\kappa}=\mu+\kappa h$ denoting the valley-dependent effective chemical potential. This analysis here reveal that the interband LOCs exhibit anisotropic behavior, which will be explicitly demonstrated in the following two subsections.

\subsection{Interband LOCs for untilted case \label{SubSecB}}

\begin{figure*}[htbp]
	\centering
	\includegraphics[width=18cm]{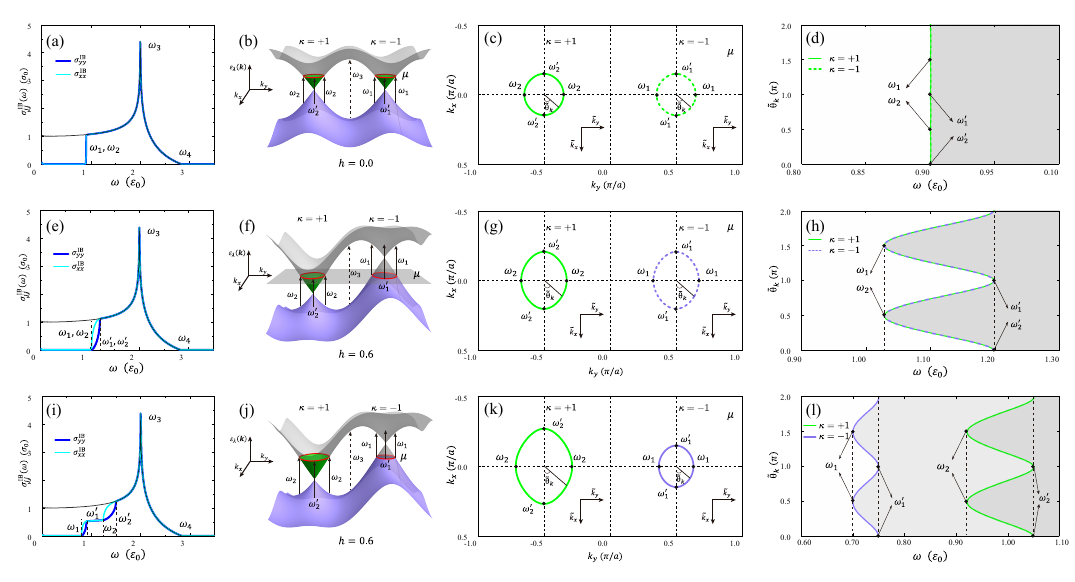}
	\caption{Results and analysis for untilted Dirac bands ($t=0$): (a,e,i) interband LOCs, (b,f,j) interband optical transitions, (c,g,k) Fermi surfaces, and (d,h,l) lower boundaries of incident photon frequency $\omega$ for arbitrary direction of wave vector $\tilde{\theta}_{k}$. The subscript $j$ in $\omega_{j}$, $\omega_{j}^{\pm}$ and $\omega_{j}^{\prime}$ are labeled as $j=1$ for $\kappa=-$ and $j=2$ for $\kappa=+1$. Panels (a)-(d) are for $h=0$ and $\mu=0.45~\varepsilon_{0}$, (e)-(h) are for $h>0$ and $\mu=0$, and (i)-(l) are for $h>0$ and $\mu=0.15~\varepsilon_{0}$. The black lines in panels (a), (e), and (i) denote the interband LOCs for $t=0$, $h=0$, and $\mu=0$, which is taken as a reference. The sharp-peak frequency $\omega_3$ and the cutoff frequency $\omega_4$ are two robust critical frequencies. Panels (d), (h), and (l) are obtained by utilizing Eqs. (\ref{EqConstrain}) and (\ref{EqTransition}), in which the shaded region contributes to the interband LOCs. }
	\label{fig2}
\end{figure*}

For the untilted case ($t=0$), the interband optical transitions are only characterized by the values of both doping $\mu$ and shifting $h$. As shown in Figs.~\ref{fig2}(a-d), when two Dirac points are unshifted ($h=0$), the two Fermi surfaces with respect to the corresponding Dirac points for the $n$-doped case ($\mu>0$) in the untilted energy bands are degenerate, leading to two degenerate conventional critical frequencies denoted as $\omega_{1}$ and $\omega_{2}$, namely, $\omega_{1}=\omega_{2}$. Throughout this work, we use the subscript $j$ in $\omega_{j}$, $\omega_{j}^{\pm}$, and $\omega_{j}^{\prime}$ with the following convention: $j=1$ corresponds to $\kappa=-$, and $j=2$ corresponds to $\kappa=+1$. In accompany with these conventional critical frequencies, there are two associated degenerate partner frequencies denoted as $\omega_{1}^{\prime}$ and $\omega_{2}^{\prime}$ with $\omega_{1}^{\prime}=\omega_{2}^{\prime}$. Further, the partner frequency $\omega_{1}^{\prime}$ is equal to the conventional critical frequency $\omega_{1}$ with $\omega_{1}^{\prime}=\omega_{1}$. It is because the two Fermi surfaces are of the same Fermi wave vector along arbitrary direction that the characteristic critical frequencies
\begin{align}
\omega_{1}^{\prime}=\omega_{2}^{\prime}=\omega_{1}=\omega_{2}=2\mu,\nonumber
\end{align}
and the interband LOCs $\sigma_{xx}^{\mathrm{IB}}(\omega,\mu>0,t=0,h=0)=\sigma_{yy}^{\mathrm{IB}}(\omega,\mu>0,t=0,h=0)$. 

When the two Dirac points are shifted oppositely along the energy direction ($h>0$), the untilted energy bands and their corresponding Fermi surfaces for the undoped case ($\mu=0$) resemble those of the doped case ($\mu>0$) without Dirac-point shifting ($h=0$), as illustrated in Figs.~\ref{fig2}(b,c) and Figs.~\ref{fig2}(f,g). This similarity arises from the interplay between the finite energy shift ($h>0$) and zero chemical potential ($\mu=0$), leading to that the conventional critical frequencies $\omega_1$ and $\omega_2$ in the former case [Figs.~\ref{fig2}(f,g)] behave analogously to their counterparts in the latter case [Figs.~\ref{fig2}(b,c)]. 

For $h\varepsilon_{0}+(-1)^{j}\mu\ge 0$, as clearly shown in Figs.~\ref{fig2}(g) and \ref{fig2}(h), the conventional critical frequencies
\begin{align}
	\omega_{1}&=\omega_{2}=\frac{2h\varepsilon_{0}}{\sqrt{1+h^2}}\nonumber
\end{align}
appear at $\tilde{\theta}_{k}=\pi/2$ and $3\pi/2$, while their partner frequencies
\begin{align}
	\omega_{1}^{\prime}&=\omega_{2}^{\prime}=2h\varepsilon_0>\omega_{1}\nonumber
\end{align}
occur at $\tilde{\theta}_{k}=0$ and $\pi$. The partner frequency $\omega_{1}^{\prime}$ differs from the conventional critical frequency $\omega_{1}$ because the Fermi wave vector along the $k_x$-direction is distinct from that along the $k_y$-direction. As evidently shown in Fig.\ref{fig2}(e), the interband LOCs $\sigma_{xx}^{\mathrm{IB}}(\omega,\mu>0,t=0,h=0)$ are exactly equal to $\sigma_{yy}^{\mathrm{IB}}(\omega,\mu>0,t=0,h=0)$ in the regions $\omega<\omega_{1}$ or $\omega>\omega_{1}^{\prime}$, but differ from the latter in the interval $\omega_{1}<\omega<\omega_{1}^{\prime}$. This result indicates that a finite energy shift ($h>0$) introduces a new characteristic frequency $\omega_{j}^{\prime}$ and leads to notable anisotropic behavior in the interband LOCs around this frequency.

For the doped case ($\mu>0$) with shifting of Dirac points ($h>0$), the physics of interband LOCs becomes richer, since the Fermi wave vectors differ not only between the two Dirac points but also along the $k_x$ and $k_y$ directions, as shown in Figs.~\ref{fig2}(j) and \ref{fig2}(k). In this scenario, the conventional critical frequency $\omega_{2}$ and its partner $\omega_{2}^{\prime}$ emerge as counterparts of $\omega_{1}$ and $\omega_{1}^{\prime}$, respectively. Unlike the two previous cases, the condition $h\varepsilon_{0}+(-1)^{j}\mu\ge 0$ is always satisfied, and the conventional critical frequencies  
\begin{align}
	\omega_{1}&
	=\frac{2\left|h\sqrt{(1+h^2)\varepsilon_{0}^{2}-\mu^2}-\mu\right|}{1+h^2},\nonumber\\
	\omega_{2}&=\frac{2\left|h\sqrt{(1+h^2)\varepsilon_{0}^{2}-\mu^2}+\mu\right|}{1+h^2}>\omega_{1},\nonumber
\end{align}
which emerge at $\tilde{\theta}_{k}=\pi/2$ and $3\pi/2$, remain non-degenerate. Similarly, the partner frequencies  
\begin{align}
	\omega_{1}^{\prime}&=2|h\varepsilon_{0}-\mu|,\nonumber\\
	\omega_{2}^{\prime}&=2|h\varepsilon_{0}+\mu|>\omega_{1}^{\prime},\nonumber
\end{align}
appearing at $\tilde{\theta}_{k}=0$ and $\pi$, are also non-degenerate. Furthermore, as seen in Figs.~\ref{fig2}(k) and \ref{fig2}(l), the distinctive behavior of the interband LOCs in the interval $\omega_{1}<\omega<\omega_{1}^{\prime}$ is replicated in the interval $\omega_{2}<\omega<\omega_{2}^{\prime}$. In summary, a finite energy shift ($h>0$) combined with a finite chemical potential ($\mu>0$) gives rise to two non-degenerate conventional critical frequencies ($\omega_{1}$ and $\omega_{2}$) and two non-degenerate partner frequencies ($\omega_1^{\prime}$ and $\omega_2^{\prime}$), leading to correspondingly rich anisotropic signatures in the interband LOCs around these frequencies. As clearly illustrated in Fig.\ref{fig2}(i), the interband LOC $\sigma_{xx}^{\mathrm{IB}}(\omega,\mu>0,t=0,h=0)$ exactly equals $\sigma_{yy}^{\mathrm{IB}}(\omega,\mu>0,t=0,h=0)$ in the regions $\omega<\omega_{1}$, $\omega_1^{\prime}<\omega<\omega_{2}$, and $\omega>\omega_{2}^{\prime}$, but differs from the latter in the intervals $\omega_1<\omega<\omega_1^{\prime}$ and $\omega_2<\omega<\omega_2^{\prime}$.

\begin{figure*}[htbp]
	\centering
	\includegraphics[width=18cm]{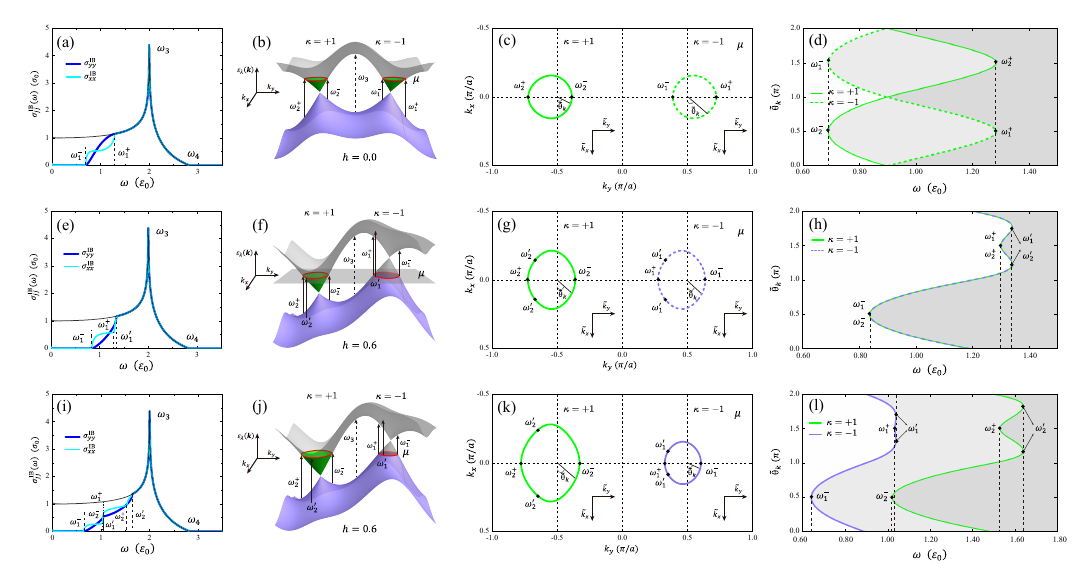}
	\caption{Results and analysis for under-tilted Dirac bands ($0<t<1$). (a,e,i) interband LOCs, (b,f,j) interband optical transitions, (c,g,k) Fermi surfaces, and (d,h,l) lower boundaries of incident photon frequency $\omega$ for an arbitrary wave-vector direction $\theta_{k}$. The subscript $j$ in $\omega_{j}$, $\omega_{j}^{\pm}$, and $\omega_{j}^{\prime}$ is labeled as $j=1$ for $\kappa=-$ and $j=2$ for $\kappa=+1$. Panels (a)–(d) correspond to $h=0$ and $\mu=0.45\,\varepsilon_{0}$; panels (e)–(h) to $h>0$ and $\mu=0$; and panels (i)–(l) to $h>0$ and $\mu=0.15\,\varepsilon_{0}$. The black curves in (a), (e), and (i) represent the interband LOCs for $t=0$, $h=0$, and $\mu=0$, which serve as a reference. The sharp-peak frequency $\omega_3$ and the cutoff frequency $\omega_4$ are two robust critical frequencies. The tilting parameter is set to $t=0.3$ for illustration. Panels (d), (h), and (l) are obtained using Eqs. (\ref{EqConstrain}) and (\ref{EqTransition}), with the shaded regions indicating the contributions to the interband LOCs.}
	\label{fig3}
\end{figure*}

\subsection{Interband LOCs for under-tilted case\label{SubSecC}}

Compared to the untilted case ($t=0$), the key distinctions in the under-tilted case ($0<t<1$) lie in the tilted energy bands and the corresponding Fermi surfaces around the Dirac points, which exhibit two distinct Fermi wave vectors along the $k_y$-direction, as clearly illustrated in Fig.~\ref{fig2} and Fig.~\ref{fig3}. Consequently, each conventional critical frequency $\omega_{j}$ (with $j=1,2$) splits into two non-degenerate conventional critical frequencies denoted as $\omega_{j}^{+}$ and $\omega_{j}^{-}$, where $\omega_{j}^{+} > \omega_{j}^{-}$. These correspond to interband optical transitions at the maximum and minimum of the Fermi wave vector along the $k_y$-direction, as shown in Figs.~\ref{fig3}(c,d), \ref{fig3}(g,h), and \ref{fig3}(k,l).

When the Dirac points are unshifted ($h=0$) in the $n$-doped case ($\mu>0$), as seen from Figs.~\ref{fig3}(a-d), the two non-degenerate conventional critical frequencies $\omega_{j}^{-}$ and $\omega_{j}^{+}$ satisfy the relations  
\begin{align}
	\omega_{1}^{-}&=\omega_{2}^{-}=\frac{2\mu}{\left|1+t\right|},\nonumber\\
	\omega_{1}^{+}&=\omega_{2}^{+}=\frac{2\mu}{\left|1-t\right|}>\omega_{1}^{-},\nonumber
\end{align}
with $\omega_{1}^{+}$ and $\omega_{2}^{-}$ appearing at $\tilde{\theta}_{k}=\pi/2$, and $\omega_{1}^{-}$ and $\omega_{2}^{+}$ occurring at $\tilde{\theta}_{k}=3\pi/2$. Notably, no partner frequency emerges in this case, as evidently shown in Figs.~\ref{fig3}(a) and \ref{fig3}(d). Fig.~\ref{fig3}(a) clearly shows that the interband LOC $\sigma_{xx}^{\mathrm{IB}}(\omega,\mu>0,0<t<1,h=0)$ equals $\sigma_{yy}^{\mathrm{IB}}(\omega,\mu>0,0<t<1,h=0)$ in the regions $\omega<\omega_{1}^{-}$ or $\omega>\omega_{1}^{+}$, but differs from it otherwise.

When the Dirac points are shifted ($h>0$) in the undoped case ($\mu=0$), Figs.~\ref{fig3}(e-h) reveal that the two non-degenerate conventional critical frequencies become
\begin{align}
	\omega_{1}^{-}&=\omega_{2}^{-}=\frac{2h }{\sqrt{h^2+(1+t)^2}}\varepsilon_{0},\nonumber\\
	\omega_{1}^{+}&=\omega_{2}^{+}=\frac{2h }{\sqrt{h^2+(1-t)^2}}\varepsilon_{0}>\omega_{1}^{-},\nonumber
\end{align}
where $\omega_{1}^{+}$ and $\omega_{2}^{-}$ emerge at $\tilde{\theta}_{k}=\pi/2$, while $\omega_{1}^{-}$ and $\omega_{2}^{+}$ occur at $\tilde{\theta}_{k}=3\pi/2$. Moreover, the Fermi surfaces associated with the two Dirac points remain degenerate in the under-tilted Dirac bands, leading again to the partner frequency
\begin{align}
	\omega_{1}^{\prime}&=\omega_{2}^{\prime}=2\sqrt{t^2+h^2}\varepsilon_{0}>\omega_{1}^{+}>\omega_{1}^{-},\nonumber
\end{align}
which appears at  
	\begin{align}
		\tilde{\theta}_{k}&=\frac{3}{2}\pi\pm\arctan\!\left[
		\frac{\arcsin\!\left(\frac{t}{\sqrt{h^2+t^2}}\right)}
		{\arcsin\!\left[\sqrt{\left(\sqrt{h^2+t^2}\right)^2-\frac{t^2}{h^2+t^2}}\,\right]}\right],\nonumber
	\end{align}
as shown in Figs.~\ref{fig3}(g) and \ref{fig3}(h). Unlike the $n$-doped case ($\mu>0$) with unshifted Dirac points ($h=0$), the partner frequency $\omega_{1}^{\prime}$ exceeds both $\omega_{1}^{-}$ and $\omega_{1}^{+}$, i.e., $\omega_{1}^{\prime}>\max\{\omega_{1}^{-},\omega_{1}^{+}\}$. Consequently, the interband LOC $\sigma_{xx}^{\mathrm{IB}}(\omega,\mu=0,0<t<1,h>0)$ equals $\sigma_{yy}^{\mathrm{IB}}(\omega,\mu=0,0<t<1,h>0)$ in the regions $\omega<\omega_{1}^{-}$ or $\omega>\omega_{1}^{\prime}$, generally differing from the latter in the intermediate frequency range.

For the case with $\mu>0$ and $h>0$, the physics of interband LOCs become more exciting due to the interplay among the band titling, chemical potential, and finite shifting along energy direction. As shown in Figs.~\ref{fig3}(i)-\ref{fig3}(l), the two Fermi surfaces with respect to the Dirac points in the under-tilted energy bands are not degenerate any longer, leading to that two non-degenerate conventional critical frequencies $\omega_{j}^{-}$ and $\omega_{j}^{+}$ satisfy the relations $\omega_{1}^{-}\neq\omega_{2}^{-}$ and $\omega_{1}^{+}\neq\omega_{2}^{+}$. Different from two previous cases, the two partner frequencies are non-degenerate, namely, $\omega_{1}^{\prime}\neq\omega_{2}^{\prime}$. Besides, the partner frequency $\omega_{j}^{\prime}$ is also greater than the maximum of $\omega_{j}^{-}$ and $\omega_{j}^{+}$, namely, $\omega_{j}^{\prime}>\mathrm{Max}\{\omega_{j}^{-},\omega_{j}^{+}\}$. As a consequence, the interband LOC $\sigma_{xx}^{\mathrm{IB}}(\omega,\mu\ge0,t=0,h=0)$ is equal to $\sigma_{yy}^{\mathrm{IB}}(\omega,\mu\ge0,t=0,h=0)$ only when $\omega$ is either greater than $\omega_{2}^{\prime}$ or less than $\omega_{1}^{-}$. Explicitly, for $\sqrt{h^2+t^{2}}\varepsilon_{0}+(-1)^{j}\mu\ge 0$, the conventional critical frequencies
\begin{align}
	\omega_{j}^{\pm}&=\frac{2\left|h^2\varepsilon_{0}^2-\mu^2\right|}{\left|h\sqrt{\left(1\mp t\right)^{2}\varepsilon_{0}^2+\left(h^2\varepsilon_{0}^2-\mu^2\right)}-(-1)^{j}\mu\left(1\mp t\right)\right|}\nonumber
\end{align}
appear at $\tilde{\theta}_{k}=\pi/2,3\pi/2$, and the partner frequencies 
\begin{align}
	\omega_{j}^{\prime}&=2\left|\sqrt{t^{2}+h^{2}}\varepsilon_{0}+(-1)^{j}\mu\right|\nonumber
\end{align}
emerge at
\begin{align}
	\tilde{\theta}_{k}&=\frac{3}{2}\pi\pm\tilde{\phi}_{j},\nonumber
\end{align}
where $\tilde{\phi}_{j}$ is defined as
\begin{align}
	\tilde{\phi}_{j}=\arctan\left[
	\frac{\arcsin \left(\frac{t}{\sqrt{h^2+t^2}}\right)}{\arcsin \left[\sqrt{\left[\sqrt{h^2+t^2}+(-1)^{j}\mu\right]^2-\frac{t^2}{h^2+t^2}}\right]}\right].\nonumber
\end{align}

	\subsection{Analytical expressions of critical frequencies \label{SubSecD}}

In this subsection, we discuss in detail the analytical expressions of the four kinds of characteristic critical frequencies appearing in the interband LOCs. First, we focus on the conventional critical frequencies and their associated partner frequencies, which depend on the Fermi surface shaped by the band tilting $t$, energy shift $h$, and chemical potential $\mu$. The analytical expressions for $\omega_{j}$ (or $\omega_{j}^{\pm}$ in under-tilted bands) and $\omega_{j}^{\prime}$ can be obtained using the Lagrange multiplier method, i.e., by optimizing the function
\begin{align}
	\mathcal{L}=2\tilde{\mathcal{Z}}\left(\tilde{k}_x,\tilde{k}_y\right)\varepsilon_{0}
	+\zeta\left[\tilde{\varepsilon}_{\lambda}^{\kappa}\left(\tilde{k}_x,\tilde{k}_y\right)-\mu\right],
\end{align}
where $\zeta$ denotes the Lagrange multiplier (see the Supplemental Materials \cite{SupplementalMaterials}).

When $t=0$, $h=0$, and $\mu>0$, for $\varepsilon_{0}+(-1)^{j}\mu\ge 0$, the conventional critical frequencies 
\begin{align}
	\omega_{j}^{\pm}&=2\mu,
\end{align}
appear at $\tilde{\theta}_{k}=\pi/2$ and $3\pi/2$,
and the partner frequencies
\begin{align}
	\omega_{j}^{\prime}&=2\mu,
\end{align}
occur at arbitrary polar angle $\tilde{\theta}_{k}$. When $t=0$, $h>0$, and $\mu\ge 0$, and under the condition $h\varepsilon_{0}+(-1)^{j}\mu\ge 0$, the conventional critical frequencies are given by
\begin{align}
	\omega_{j}^{\pm}&=\frac{2\left|h\sqrt{(1+h^2)\varepsilon_{0}^2-\mu^2}+(-1)^{j}\mu\right|}{1+h^2},
\end{align}
which emerge at $\tilde{\theta}_{k}=\pi/2$ and $3\pi/2$, while the partner frequencies take the form
\begin{align}
	\omega{j}^{\prime}&=2\left|h\varepsilon_{0}+(-1)^{j}\mu\right|,
\end{align}
and occur at $\tilde{\theta}_{k}=0$ and $\pi$. When $0<t<1$, $h=0$, and $\mu>0$, under the condition $t\varepsilon_{0}+(-1)^{j}\mu\ge 0$,  
the conventional critical frequencies $\omega_{j}^{\pm}$ are given by
\begin{align}
	\omega_{j}^{\pm}&=\frac{2\mu}{|1\mp t|},
\end{align}
and emerge at $\tilde{\theta}_{k}=\pi/2$ and $3\pi/2$,  
while no partner frequency appears in this case.

When $0<t<1$, $h>0$, and $\mu\ge0$, under the condition $\sqrt{h^2+t^{2}}\,\varepsilon_{0}+(-1)^{j}\mu\ge 0$, the conventional critical frequencies are given by 
\begin{align}
	\omega_{j}^{\pm}&=\frac{2\left|h^2\varepsilon_{0}^2-\mu^2\right|}{\left|h\sqrt{\left(1\mp t\right)^{2}\varepsilon_{0}^2+\left(h^2\varepsilon_{0}^2-\mu^2\right)}-(-1)^{j}\mu\left(1\mp t\right)\right|},
\end{align}
and emerge at $\tilde{\theta}_{k}=\pi/2$ and $3\pi/2$, while the partner frequencies take the form
\begin{align}
	\omega_{j}^{\prime}&=2\left|\sqrt{t^{2}+h^{2}}\,\varepsilon_{0}+(-1)^{j}\mu\right|,
\end{align}
and occur at
\begin{align}
	\tilde{\theta}_{k}&=\frac{3}{2}\pi\pm \tilde{\phi}_{j},
\end{align}
where
\begin{align}
	\tilde{\phi}_{j}&=\arctan\!\left[
	\frac{\arcsin\!\left[\frac{t}{\sqrt{h^2+t^2}}\right]}{\arcsin\!\left[\sqrt{\left(\sqrt{h^2+t^2}+(-1)^{j}\mu\right)^2-\frac{t^2}{h^2+t^2}}\,\right]}\right],\nonumber
\end{align}
which are determined by the conditions
\begin{align}
	&\sin \left(a\tilde{k}_{y}\right)=\sin \left(a\tilde{k} \sin\tilde{\theta}_{k}\right)
	\nonumber\\&
	=-\frac{t}{\sqrt{h^2+t^2}},\\
	&\sin \left(a\tilde{k}_{x}\right)=\sin \left(a\tilde{k} \cos\tilde{\theta}_{k}\right)
	\nonumber\\&
	=\pm\sqrt{\left(\sqrt{h^2+t^2}+(-1)^{j}\frac{\mu}{\varepsilon_{0}}\right)^2-\frac{t^2}{h^2+t^2}}.
\end{align}

We emphasize that the above analytical expressions for the conventional critical frequencies $\omega_{j}$ (or $\omega_{j}^{\pm}$), the partner frequencies $\omega_{j}^{\prime}$, and the corresponding polar angles $\tilde{\theta}_{k}$ provide a quantitative account of the results obtained from numerical calculations (see the Supplemental Materials \cite{SupplementalMaterials}).

\begin{figure}[htbp]
	\centering
	\includegraphics[width=8cm]{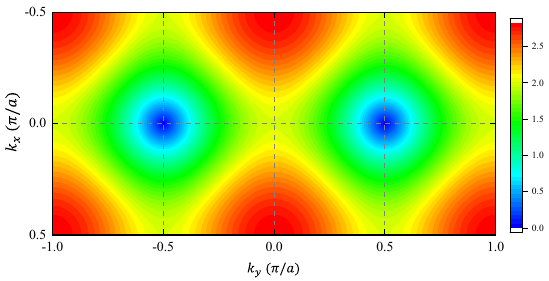}
	\caption{Density plot of the interband optical transition determined by $\omega=2\mathcal{Z}(k_x,k_y)\varepsilon_{0}$. The sharp-peak frequency $\omega_{3}=2~\varepsilon_{0}$ denotes the maximum of frequency for interband optical transition at seven high-symmetry points $(0,0)$, $(0,\pm \frac{\pi}{a})$, $(\pm \frac{\pi}{2a},\pm \frac{\pi}{2a})$. The cutoff frequency $\omega_4=2\sqrt{2}~\varepsilon_{0}$ measures the maximum of interband optical transitions between the lowest energy and the highest energy at six high-symmetry points $(\pm \frac{\pi}{2a},0)$, and $(\pm \frac{\pi}{2a},\pm \frac{\pi}{a})$.}
	\label{fig4}
\end{figure}

Next, we turn to the sharp-peak frequency $\omega_3$ and the cutoff frequency $\omega_4$, which are determined solely by the energy bands and are independent of the Fermi surface. Using the relation
\begin{align}
	\omega=\varepsilon_{+}(k_x,k_y)-\varepsilon_{-}(k_x,k_y)=2\mathcal{Z}(k_x,k_y)\varepsilon_{0},
	\label{IOTdiff}
\end{align}
we present the density plot of $\omega=2\mathcal{Z}(k_x,k_y)\varepsilon_{0}$ in Fig.~\ref{fig4}. The interband LOCs display a sharp peak at $\omega_{3}$, which corresponds to the maximum frequency for interband optical transitions at seven high-symmetry points in the $k_x$-$k_y$ plane: $(0,0)$, $(0,\pm \frac{\pi}{a})$, and $(\pm \frac{\pi}{2a},\pm \frac{\pi}{2a})$. This yields the value $\omega_{3}=2\varepsilon_{0}$. The sharp peak originates from van Hove singularities at $\omega=\omega_3$. Beyond the sharp-peak frequency $\omega=\omega_3$, the interband LOCs are gradually suppressed and eventually vanish at the cutoff frequency $\omega=\omega_4=2\sqrt{2}\,\varepsilon_{0}$. It is noted that $\omega_4$ measures the maximum of energy in the interband transition between the lowest and highest energies at six high-symmetry points: $(\pm \frac{\pi}{2a},0)$ and $(\pm \frac{\pi}{2a},\pm \frac{\pi}{a})$, as a consequence of the Pauli exclusion principle and the finite boundary of the Brillouin zone. We emphasize that, as illustrated in Fig.~\ref{fig4}, both the sharp-peak frequency $\omega_3$ and the cutoff frequency $\omega_4$ are robust critical frequencies, independent of $\mu$, $t$, and $h$---a conclusion further supported by Figs.~\ref{fig2}(a,e,i) and Figs.~\ref{fig3}(a,e,i).

\section{Comparisons with the linearized $k\cdot p$ Hamiltonian \label{Sec:comparisons}}

To highlight the characteristic properties of interband LOCs in 2D tilted Dirac bands, we compare the results obtained from the TB Hamiltonian and the linearized $k \cdot p$ Hamiltonian. In the vicinity of the two Dirac points $(0,\kappa\frac{\pi}{2a})$, the TB Hamiltonian in Eq.~(\ref{Eq1}) reduces to the linearized $k \cdot p$ Hamiltonian describing a pair of oppositely tilted Dirac fermions. The resulting linearized $k \cdot p$ Hamiltonian and its eigenvalue are given by
\begin{align}
	\mathscr{H}_\kappa(\tilde{\boldsymbol{k}})
	=\mathscr{H}_\kappa(\tilde{k}_{x},\tilde{k}_{y})
	&=\kappa (t a \tilde{k}_y - h)\varepsilon_0\tau_0
	\nonumber\\&
	+ a \tilde{k}_x \varepsilon_0 \tau_1 + \kappa a \tilde{k}_y\varepsilon_0 \tau_2,
\end{align}
and
\begin{align}
	\mathscr{E}_\kappa^\lambda(\tilde{k}_x,\tilde{k}_y)=\kappa (t a \tilde{k}_y- h)\varepsilon_0  +\lambda a\mathscr{Z}(\tilde{k}_x,\tilde{k}_y) \varepsilon_0,
	\label{Eq2}
\end{align}
where $\mathscr{Z}(\tilde{k}_x,\tilde{k}_y)=\sqrt{\tilde{k}_x^2+\tilde{k}_y^2}=|\tilde{\boldsymbol{k}}|$. 
\\
\begin{figure*}[htbp]
	\centering
	\includegraphics[width=18cm]{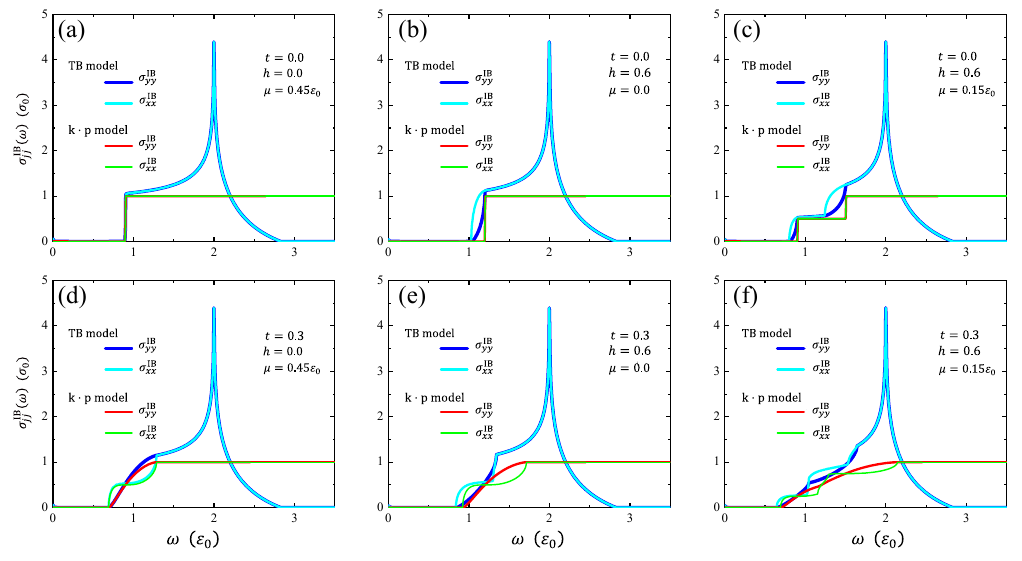}
	\caption{Comparison of interband LOCs between the linearized $k \cdot p$ model and the TB model for tilted Dirac bands. The analytical expressions for the interband LOCs in the linearized $k \cdot p$ model are taken from Ref. \cite{PRBTan2022}.}
	\label{fig5}
\end{figure*}

The Matsubara Green's function is given by
\begin{align}
	\mathscr{G}_{\kappa}(\tilde{\boldsymbol{k}},i\Omega_{m})&=\left[(i\Omega_m+\mu)\tau_0-\mathscr{H}_\kappa(\tilde{\boldsymbol{k}})\right]^{-1}
	\nonumber\\&
	=\frac{1}{2}\sum_{\lambda=\pm} \frac{\mathscr{P}_\kappa^\lambda(\tilde{\boldsymbol{k}})}{i\Omega_m+\mu-\mathscr{E}_\kappa^\lambda(\tilde{k}_x,\tilde{k}_y)},
\end{align}
with
\begin{align}
	\mathscr{P}_\kappa^\lambda(\boldsymbol{k})&=\tau_0 +\lambda\frac{\tilde{k}_x\tau_1+\kappa \tilde{k}_y\tau_2}{\mathscr{Z}(\tilde{k}_x,\tilde{k}_y)}.
\end{align}

Introducing the electromagnetic field via minimal coupling and expanding the Hamiltonian to the leading order of $e$, we obtain
\begin{align}
	\mathscr{H}_{\kappa}(\tilde{\boldsymbol{k}}+e\boldsymbol{A})
	&=\mathscr{H}_{\kappa}(\tilde{\boldsymbol{k}})
	+ea\varepsilon_{0}\tau_1A_x+\kappa ea\varepsilon_{0}(t\tau_0+\tau_2).\nonumber
\end{align}
Consequently, the current operators $\hat{\mathscr{J}}_j^{\kappa}$ with $j=x,y$ are given by
\begin{align}
	\hat{\mathscr{J}}_x^{\kappa}&=\frac{\partial \mathscr{H}_{\kappa}(\tilde{\boldsymbol{k}}+e\boldsymbol{A})}{\partial A_x}
	=ea\varepsilon_{0}\tau_1, \label{current_x}\\
	\hat{\mathscr{J}}_y^{\kappa}&=\frac{\partial \mathscr{H}_{\kappa}(\tilde{\boldsymbol{k}}+e\boldsymbol{A})}{\partial A_y}
	=\kappa ea\varepsilon_{0}(t\tau_0+\tau_2).
	\label{current_y}
\end{align}
The current-current correlation function is defined as
\begin{align}
	\Pi_{ij}(\omega,\mu,t,h)
	&=\sum_{\kappa=\pm}\Pi_{ij}^{\kappa}(\omega,\mu,t,h),
\end{align}
with
\begin{align}
	&\Pi_{ij}^{\kappa}(\omega,\mu,t,h)
	=\lim_{\boldsymbol{q}\to 0}\frac{1}{\beta}\sum_{i\Omega_m}
	\int_{-\infty}^{+\infty}\int_{-\infty}^{+\infty}\frac{\mathrm{d}^2\tilde{\boldsymbol{k}}}{(2\pi)^2}
	\nonumber\\&
	\mathrm{Tr}\left[\hat{\mathscr{J}}_i^{\kappa} \mathscr{G}_{\kappa}(\tilde{\boldsymbol{k}},i\Omega_{m}) \hat{\mathscr{J}}_j^{\kappa} \mathscr{G}_{\kappa}(\tilde{\boldsymbol{k}}+\boldsymbol{q},i\Omega_{m}+\omega+i\eta)\right].
\end{align}

To elucidate the relationship between the LOCs and the TB energy bands, we focus on the interband optical transitions between the valence and conduction bands. After straightforward algebraic manipulation, the real part of the interband LOCs can be written as
\begin{align}
	\mathrm{Re}~\sigma_{jj}^{\mathrm{IB}}&(\omega,\mu,t,h)
	=e^2 \pi \sum_{\kappa=\pm}\int_{-\frac{\pi}{2a}}^{+\frac{\pi}{2a}} \frac{\mathrm{d} \tilde{k}_x}{2\pi} \int_{-\frac{\pi}{2a}}^{+\frac{\pi}{2a}} \frac{\mathrm{d} \tilde{k}_y}{2\pi} 
	\nonumber\\&\times
	\mathcal{F}_{\kappa;jj}^{-,+}(\tilde{k}_x,\tilde{k}_y)\,
	\delta\!\left[\omega-2a\mathscr{Z}(\tilde{k}_x,\tilde{k}_y) \varepsilon_0\right]
	\nonumber\\&\times
	\frac{f\!\left[\mathscr{E}_{\kappa}^{-}(\tilde{k}_x,\tilde{k}_y)\right]-f\!\left[\mathscr{E}_{\kappa}^{+}(\tilde{k}_x,\tilde{k}_y)\right]} {\omega},
	\label{integrationReg2}
\end{align}
where
\begin{align}
	\mathcal{F}_{\kappa;xx}^{-,+}(\tilde{k}_x,\tilde{k}_y)
	&=(a\varepsilon_0)^2\frac{\tilde{k}_y^2}{\tilde{k}_x^2+\tilde{k}_y^2}, \\
	\mathcal{F}_{\kappa;yy}^{-,+}(\tilde{k}_x,\tilde{k}_y)
	&=(a\varepsilon_0)^2\frac{\tilde{k}_x^2}{\tilde{k}_x^2+\tilde{k}_y^2}.
\end{align}

By setting $h=0$, $a\varepsilon_0= \hbar v_F$, and $t a\varepsilon_0=t\hbar v_F=\hbar v_t$, the Hamiltonian in Eq.~(\ref{Eq2}) reduces to
\begin{align}
	\mathscr{H}_\kappa(\tilde{k}_x,\tilde{k}_y)=\kappa \hbar v_t \tilde{k}_y\tau_0+ \hbar v_F\tilde{k}_x\tau_1+\kappa \hbar v_F\tilde{k}_y\tau_2,
\end{align}
which corresponds to the linearized $k\cdot p$ Hamiltonian \cite{PRBTan2022} in the isotropic limit ($v_x=v_y= v_F$). In the following, we compare the interband LOCs obtained from the TB model with the analytical results derived from the linearized $k\cdot p$ Hamiltonian \cite{PRBTan2022}. For the comparisons presented in Fig.~\ref{fig5}, we adopt two convenient approximations. First, to incorporate the energy-shifting of the Dirac point, the chemical potential in the analytical expressions of Ref.~\cite{PRBTan2022} is replaced by the valley-dependent effective chemical potentials $\mu_{\kappa}=\mu+\kappa h$. This replacement is natural because the chemical potential is measured relative to the corresponding Dirac point, thereby allowing the analytical formulas to be extended to the case of a nonzero energy-shifting. Second, the finite integration limits $\pm \frac{\pi}{2a}$ in Eq.~(\ref{integrationReg2}) can be safely extended to $\pm \infty$, since the Fermi-Dirac distribution decays rapidly at the low temperature assumed in our numerical calculations ($T=1~\mathrm{K}$). Hence, contributions from the regions $(-\infty,-\frac{\pi}{2a})$ and $(+\frac{\pi}{2a},+\infty)$ in integrating over both $\tilde{k}_{x}$ and $\tilde{k}_{y}$ are negligible. Consequently, the analytical expressions from Ref.~\cite{PRBTan2022} provide an excellent approximation to the integral in Eq.~(\ref{integrationReg2}) and can be directly compared with the numerical results obtained from Eq.~(\ref{integrationReg1}) in the TB model.

As shown in Fig.~\ref{fig5}, the conventional critical frequencies $\omega_{j}$ (or $\omega_{j}^{\pm}$ in the under-tilted bands) and the interband LOCs in the region $0<\omega<\max\{\omega_{1}^{+},\omega_{2}^{+}\}$ behave qualitatively similar to those obtained from the linearized $k\cdot p$ model \cite{PRBTan2022}. However, the partner frequencies $\omega_{j}^{\prime}$ appear only in the TB model. More importantly, the sharp-peak frequency $\omega_3$ and the cutoff frequency $\omega_4$, which are absent in the linearized $k\cdot p$ model, emerge as distinctive features of the TB description. Furthermore, in the regions where $\omega>\max\{\omega_{1}^{+},\omega_{2}^{+},\omega_{1}^{\prime},\omega_{2}^{\prime}\}$, the behavior of the interband LOCs in the TB model differs significantly from the step-like profiles predicted by the linearized $k\cdot p$ model. These comparisons demonstrate that the linearized $k\cdot p$ model does not always provide an adequate description of the optical properties in 2D Dirac bands.

\section{Conclusions and Discussions \label{Sec:conclusions and discussions}}

Within the linear response theory, we theoretically investigated the interband LOCs in 2D tilted Dirac bands using a TB model that incorporates both band tilting and Dirac-point shifting. We identified four characteristic types of critical frequencies in the interband LOCs of the TB model: the conventional critical frequencies, the partner frequencies, the sharp-peak frequency, and the cutoff frequency. The latter three types are consistently absent in the corresponding linearized $k\cdot p$ model. The origins of these characteristic frequencies were clarified through analytical expressions derived either via the Lagrange multiplier method or by analyzing interband optical transitions at high-symmetry points of the energy bands. Our comparisons of the interband LOCs demonstrate that the linearized $k\cdot p$ model is not always sufficient to capture the optical properties of 2D Dirac systems. The theoretical predictions presented here can guide future experimental studies of tilt-dependent phenomena in the optical response of 2D tilted Dirac materials.

We highlight three pertinent issues for further consideration. The first concerns the extension of the TB framework to encompass gapped tilted Dirac bands—for instance, by adding a gap term $\Delta_{g}\tau_{3}$ to the TB Hamiltonian in Eq.~(\ref{Eq1})—as well as its generalization to non-linearized low-energy $k \cdot p$ Hamiltonians with a finite gap, such as that of 1$T^{\prime}$-MoS$_{2}$. To describe the valley-spin-polarized energy bands of 1$T^{\prime}$-MoS$_{2}$ under a vertical electric field, the low-energy $k \cdot p$ Hamiltonian \cite{HRC2014}
	\begin{align}
		\mathcal{H}(k_x,k_y)&
		=\frac{(\tau_0+\tau_{3})\otimes\sigma_0}{2}E_p(k_x,k_y)
		\nonumber\\&
		+\frac{(\tau_0-\tau_{3})\otimes\sigma_0}{2}E_d(k_x,k_y)
		\nonumber\\&
		+\left(\tau_{2}\otimes\sigma_0\right)\hbar v_{1}k_{x}
		+\left(\tau_{1}\otimes\sigma_{1}\right)\hbar v_2k_y
		\nonumber\\&
		+\left(\tau_{1}\otimes\sigma_0\right)\left|\frac{E_{z}}{E_c}\right| \hbar v_2 \Lambda,
	\end{align}
	was originally proposed in early January 2015 by one of the present authors through the inclusion of the final term, $\left(\tau_{1}\otimes\sigma_0\right)\left|E_{z}/E_c\right| \hbar v_2 \Lambda$, into the $k \cdot p$ Hamiltonian presented in Ref. \cite{Science2014}. In this modified low-energy $k \cdot p$ Hamiltonian \cite{Science2014,HRC2014}, $E_p(k_x,k_y)=-\delta-\frac{\hbar^{2} k_{x}^{2}}{2m_{x}^{p}}-\frac{\hbar^{2} k_{y}^{2}}{2m_{y}^{p}}+B$, $E_d(k_x,k_y)=\delta+\frac{\hbar^{2} k_{x}^{2}}{2m_{x}^{d}}+\frac{\hbar^{2} k_{y}^{2}}{2m_{y}^{d}}+B$;
	$\tau$ and $\sigma$ are Pauli matrices acting on the orbital ($p$-and $d$-orbital) and spin spaces, respectively; $E_{z}$ denotes the vertical electric field and $E_c$ is its critical value. The points $(0,\kappa \Lambda)$ represent the intersections of $E_{d}(k_x,k_y)$ and $E_{p}(k_x,k_y)$, with $\kappa=\pm$ and $\Lambda=0.139\times 10^{10}~\mathrm{m}^{-1}$. The model parameters for 1$T^{\prime}$-MoS$_{2}$ obtained by fitting to first-principles band structures, are:
	$\delta=-0.33~\mathrm{eV}$, $v_1=3.87\times10^5~\mathrm{m/s}$, $v_2=0.46\times10^5~\mathrm{m/s}$, $m_{x}^{p}=0.50~m_{e}$, $m_{y}^{p}=0.16~m_{e}$, $m_{x}^{d}=2.48~m_{e}$, $m_{y}^{d}=0.37~m_{e}$, where $\delta<0$ corresponds to a $d$-$p$ band inversion, and $m_e$ is the free electron mass. Specifically, the parameter $B=0.13~\mathrm{eV}$ describes an overall energy shift. In the vicinity of the Dirac point at $(0,\kappa \Lambda)$, the linearized $k \cdot p$ Hamiltonian \cite{PRBTan2021}
	\begin{align}
		\mathscr{H}_{\kappa}(\tilde{k}_x,\tilde{k}_y)
		&=\hbar \tilde{k}_x v_1 \gamma_{1}+\hbar \tilde{k}_y\left(v_2 \gamma_{0}-\kappa v_{-}I -\kappa v_{+}\gamma_{2}\right)
		\nonumber\\&
		+\Delta_{\mathrm{so}}\left(\kappa \gamma_{0}-i \alpha \gamma_{1}\gamma_{2}\right),
	\end{align}
	can be obtained by substituting $\tilde{k}_y+\kappa \Lambda$ for $k_y$ and retaining only terms linear in $\tilde{k}_x$ and $\tilde{k}_y$. Here $I=\tau_{0}\otimes\sigma_{0}$, $\gamma_{0}=\tau_{1}\otimes\sigma_{1}$, $\gamma_{1}=\tau_{2}\otimes\sigma_{0}$, $\gamma_{2}=\tau_{3}\otimes\sigma_{0}$, $-i\gamma_{1}\gamma_{2}=\tau_{1}\otimes\sigma_{0}$, $\Delta_{\mathrm{so}}=\hbar \Lambda v_{2}=0.042~\mathrm{eV}$, $v_{-}\equiv\frac{\hbar\Lambda}{2m_y^p}-\frac{\hbar\Lambda}{2m_y^d}=2.86\times 10^{5}~\mathrm{m/s}$, $v_{+}\equiv\frac{\hbar\Lambda}{2m_y^p}+\frac{\hbar\Lambda}{2m_y^d}=7.21\times 10^{5}~\mathrm{m/s}$, and $\alpha=\left|E_{z}/E_c\right|$. The derivation uses the two relations $-\delta-\frac{\hbar^{2} \Lambda^{2}}{2m_{y}^{p}}+B=0$ and $\delta+\frac{\hbar^{2} \Lambda^{2}}{2m_{y}^{d}}+B=0$.

In the absence of a vertical electric field, the Dirac bands and energy gaps derived from the low-energy $k \cdot p$ Hamiltonian of 1$T^{\prime}$-MoS$_{2}$ are spin-degenerate---similar to the situation shown in Fig.~\ref{fig3}(b) apart from a nonzero indirect gap---and consequently the interband LOCs are expected to resemble those in Fig.~\ref{fig3}(a). When a vertical electric field is applied, the Dirac bands and gaps obtained from the low-energy $k \cdot p$ Hamiltonian become valley-spin-polarized. As a result, the conventional characteristic frequencies $\omega_{\kappa}^{\pm}$ and the partner frequency $\omega_{\kappa}^{\prime}$ split into spin-dependent counterparts $\omega_{\kappa s}^{\pm}$ and $\omega_{\kappa s}^{\prime}$, in qualitative agreement with the predictions of its linearized $k \cdot p$ Hamiltonian \cite{PRBTan2021}. Moreover, the sharp-peak frequency $\omega_{3}$ appears in the low-energy $k \cdot p$ Hamiltonian but is absent in its linearized version \cite{PRBTan2021}. However, the cutoff frequency $\omega_{4}$ is missing in both the low-energy $k \cdot p$ Hamiltonian and its linearized counterpart. These analyses indicate that the conclusions drawn from the low-energy $k \cdot p$ Hamiltonian of 1$T^{\prime}$-MoS$_{2}$ are largely consistent with those derived from the TB Hamiltonian  except for the absence of cutoff frequency, and are similar to those from the linearized low-energy $k \cdot p$ Hamiltonian of 1$T^{\prime}$-MoS$_{2}$ for the presence of partner frequencies and sharp-peak frequency.

The second issue addresses the possible influence of energy warping and lattice anisotropy on the interband LOCs. Generally, warping effects in Dirac bands become significant when the Fermi energy is tuned far from the Dirac point \cite{PRLFu2009,PRLSouma2011,PRBNomura2014}, and have been shown to induce a variety of unique phenomena---for instance, double Andreev reflection \cite{PRBYu2017,PRBHou2017,FOP2020} and modifications to optical conductivity \cite{PRBCarbotte2013,PRBXiao2013} caused by hexagonal warping in topological insulators. It is anticipated that warping effects in tilted Dirac bands would give rise to complex angular dependence of the interband LOCs. In the parameter setting of $t=0$ and $h=0$, the TB Hamiltonian in Eq.(\ref{TB2}) exhibits circular symmetry in the low-energy regime where the Fermi energy is near the Dirac point but reduces to tetragonal symmetry (resulting in tetragonal warping) in the higher-energy regime where the Fermi energy is tuned far from the Dirac point. The tetragonal warping leads to a squarish Fermi surface, producing a fourfold periodic angular dependence of interband LOCs, unlike the straight line in Fig.\ref{fig2} (d) dictated by circular symmetry at lower Fermi energy. In the parameter setting of $t>0$ and/or $h>0$, the symmetry of the TB Hamiltonian in Eq.(\ref{TB2}) and consequently its warping are governed by the competition among $t$, $h$, and $\mu$, which precludes the formation of a simple, robust warping pattern. Consequently, the interband LOCs in tilted Dirac bands may well exhibit a complex angular dependence.

If the lattice anisotropy is taken into account, the TB Hamiltonian in Eq.(\ref{TB2}) is replaced by
\begin{align}
	\tilde{\mathcal{H}}_{\kappa}\left(\tilde{k}_x,\tilde{k}_y\right)&=
	\mathcal{H}\left(\tilde{k}_x,\tilde{k}_y-\kappa\frac{\pi}{2a}\right)
	\nonumber\\&
	=\left[\kappa t\sin(b \tilde{k}_y)-\kappa h\cos(b \tilde{k}_y)\right]\varepsilon_{0}\tau_0
	\nonumber\\&
	\hspace{+0.25cm}
	+\sin(a \tilde{k}_x)\varepsilon_{0}\tau_1 +\kappa \sin(b \tilde{k}_y)\varepsilon_{0}\tau_2,\nonumber
\end{align}
with $\tilde{k}_x\in [-\frac{\pi}{2a},+\frac{\pi}{2a}]$ and $\tilde{k}_y\in [-\frac{\pi}{2b},+\frac{\pi}{2b}]$, whose low-energy linearized $k \cdot p$ Hamiltonian
\begin{align}
	\mathscr{H}_\kappa(\tilde{k}_{x},\tilde{k}_{y})
	&=\kappa (t b \tilde{k}_y - h)\varepsilon_0\tau_0
	+ a \tilde{k}_x \varepsilon_0 \tau_1 + \kappa b \tilde{k}_y\varepsilon_0 \tau_2,\nonumber
\end{align}
processes anisotropic Fermi velocities along $\tilde{k}_y$- and $\tilde{k}_y$-direction even when the band tilting is neglected. Based on the conclusions of optical conductivities in 2D tilted Dirac bands within linearized $k \cdot p$ Hamiltonian \cite{JPSJNishine2010,PRBVerma2017,PRBIurov2018,PRBHerrera2019,PRBRostamzadeh2019,PRBTan2021,PRBTan2022,PRBHou2023,PRBMojarro2021,PRBWild2022,PRBYao2021} and the TB Hamiltonian in this work, we expect more interesting behaviors of interband LOCs after considering the lattice anisotropy.

The third issue revolves around the intricate features of interband LOCs arising when the 2D Dirac bands become critical- or over-tilted. In over-tilted Dirac bands, two non-degenerate conventional critical frequencies $\omega_{j}^{\pm}$ emerge, whereas only a single conventional critical frequency $\omega_{j}^{-}$ appears in critically tilted bands. Even for $h=0$, under certain parameter regimes the partner frequency $\omega_{j}^{\prime}$ can split into two non-degenerate partner frequencies $\omega_{j}^{+\prime}$ and $\omega_{j}^{-\prime}$. The situation becomes richer when $h>0$ due to the more complex band structure and resulting Fermi surfaces. Third, the sharp-peak frequency $\omega_3$ and the cutoff frequency $\omega_4$ remain unchanged, as they are determined solely by interband optical transitions at high-symmetry points, which are unaffected by the tilt of the bands.
The three intriguing issues outlined above warrant further investigation in future work.

\section*{ACKNOWLEDGEMENTS\label{Sec:acknowledgements}}

	We are grateful to Prof. Long Liang for helpful discussion. This work is partially supported by the National Natural Science Foundation of China under Grants No. 11547200, No. 12204329 and No. 11874273, the Natural Science Foundation of Jiangsu Province under Grant No. BK20241929, and the Research Institute of Intelligent Manufacturing Industry Technology of Sichuan Arts and Science University. We thank the High Performance Computing Center at Sichuan Normal University.

\begin{widetext}
	
	\makeatletter
	\makeatother
	
	\begin{center}
		{\textbf{Supplemental Materials to ``Interband optical conductivities in two-dimensional tilted Dirac bands revisited within the tight-binding model''}}
	\end{center}
	
	\section{Explicit expression of the longitudinal optical conductivity\label{Sec:AppendixA}}

Introducing the electromagnetic field, we have
\begin{align}
	\mathcal{H}(\boldsymbol{k}+e\boldsymbol{A})=\tau_0\left[t\cos(a k_y +aeA_y)+h\sin(a k_y+aeA_y)\right]+\tau_1 \sin(a k_x+aeA_x)+\tau_2 \cos(a k_y+aeA_y),
\end{align}
from the tight-binding Hamiltonian in Eq.(1) of main text. To the leading order of $e$, 
\begin{align}
	\cos(a k_y +aeA_y) &=\cos(ak_y)\cos(aeA_y)-\sin(ak_y)\sin(aeA_y)
	=\cos(ak_y)\left[1+\mathcal{O}(e^2)\right] -\sin(ak_y)\left[aeA_y+\mathcal{O}(e^3)\right],\nonumber\\
	\sin(a k_y +aeA_y) &=\sin(ak_y)\cos(aeA_y)+\cos(ak_y)\sin(aeA_y)
	=\sin(ak_y)\left[1+\mathcal{O}(e^2)\right] +\cos(ak_y)\left[aeA_y+\mathcal{O}(e^3)\right],\nonumber\\
	\sin(a k_x +aeA_x) &=\sin(ak_x)\cos(aeA_x)+\cos(ak_x)\sin(aeA_x)
	=\sin(ak_x)\left[1+\mathcal{O}(e^2)\right] +\cos(ak_x)\left[aeA_x+\mathcal{O}(e^3)\right].\nonumber
\end{align}
As a consequence, we arrive at
\begin{align}
	\mathcal{H}(\boldsymbol{k}+e\boldsymbol{A}) =\mathcal{H}(\boldsymbol{k})+\tau_1eaA_x \cos(a k_x)-(t\tau_0+\tau_2)eaA_y \sin(a k_y)+h\tau_0eaA_y\cos(a k_y).
	\label{1storder}
\end{align}

The Green's function in Eq.(8) of main text can be obtained as
\begin{align}
	G(\boldsymbol{k},i\Omega_m) &=\left[\tau_0(i\Omega_m+\mu)-\mathcal{H}(\boldsymbol{k})\right]^{-1}
	=\frac{\tau_0\left[i\Omega_m+\mu-t\cos(ak_y)-h\sin(ak_y)\right]+\tau_1\sin(ak_x)+\tau_2 \cos(ak_y)}{
		\left[i\Omega_m+\mu-t\cos(ak_y)-h\sin(ak_y)\right]^2-\left[\mathcal{Z}(k_x,k_y)\right]^2}
	\nonumber\\
	&=\frac{1}{2}\sum_{\lambda=\pm} \frac{1}{i\Omega_m+\mu-\varepsilon_{\lambda}(k_x,k_y)}
	\left[\tau_0+\lambda\frac{\tau_1\sin(ak_x)+\tau_2\cos(ak_y)}{\mathcal{Z}(k_x,k_y)}\right]\nonumber\\
	&=\frac{1}{2}\sum_{\lambda=\pm} \frac{\mathcal{P}_{\lambda}(\boldsymbol{k})}{i\Omega_m+\mu-\varepsilon_{\lambda}(k_x,k_y)}
\end{align}
with $\mathcal{P}_{\lambda}(\boldsymbol{k})=\tau_0+\lambda\frac{\tau_1 \sin(ak_x)+\tau_2 \cos(ak_y)}{\mathcal{Z}(k_x,k_y)}$, and
\begin{align}
	G(\boldsymbol{k},i\Omega_m+\omega+i\eta)
	&=\frac{1}{2}\sum_{\lambda^\prime=\pm} \frac{\mathcal{P}_{\lambda^\prime}(\boldsymbol{k})}{i\Omega_m+\omega+\mu-\varepsilon_{\lambda}(k_x,k_y)+i\eta}
\end{align}
with $\mathcal{P}_{\lambda^\prime}(\boldsymbol{k}) =\tau_0+\lambda^\prime\frac{\tau_1 \sin(ak_x)+\tau_2 \cos(ak_y)}{\mathcal{Z}(k_x,k_y)}$.

After summing over Matsubara frequency $\Omega_m$ and a straightforward trace calculation, we express the longitudinal current-current correlation function $\Pi_{jj}(\omega,\mu,t,h)$ as
\begin{align}
	\Pi_{jj}(\omega,\mu,t,h)= e^2\sum_{\lambda=\pm}\sum_{\lambda^{\prime}=\pm} \int_{-\frac{\pi}{2a}}^{+\frac{\pi}{2a}}\frac{\mathrm{d}k_x}{2\pi}\int_{-\frac{\pi}{a}}^{+\frac{\pi}{a}}\frac{\mathrm{d}k_y}{2\pi} \mathcal{F}_{jj}^{\lambda,\lambda^{\prime}}(k_x,k_y) \frac{f\left[\varepsilon_{\lambda}(k_x,k_y)\right]-f\left[\varepsilon_{\lambda^{\prime}}(k_x,k_y)\right]} {\omega+\varepsilon_{\lambda}(k_x,k_y)-\varepsilon_{\lambda^{\prime}}(k_x,k_y)+i\eta},
\end{align}
where $f(x)=1/\left\{1+\exp\left[(x-\mu)/(k_B T)\right]\right\}$ is the Fermi distribution function, and  the explicit expressions of $\mathcal{F}_{jj}^{\lambda,\lambda^{\prime}}(k_x,k_y)$ are given as
\begin{align}
	\mathcal{F}_{xx}^{\lambda,\lambda^{\prime}}(k_x,k_y)
	=& \frac{a^2\cos^2(a k_x)}{2} \left\{1+\lambda\lambda^{\prime} \frac{\sin^2(ak_x)-\cos^2(ak_y)}{\left[\mathcal{Z}(k_x,k_y)\right]^2}\right\}, 
	\\
	\mathcal{F}_{yy}^{\lambda,\lambda^{\prime}}(k_x,k_y)=& \frac{a^2}{2} \left\{\sin^2(ak_y)+\left[t\sin(ak_y)-h\cos(ak_y)\right]^2 +2(\lambda+\lambda^{\prime})\frac{\sin(ak_y)\cos(ak_y)\left[t\sin(ak_y)-h\cos(ak_y)\right]}{\mathcal{Z}(k_x,k_y)} \right.\nonumber\\
	&\left.+\lambda\lambda^{\prime} \left[\left[t\sin(ak_y)-h\cos(ak_y)\right]^2-\sin^2(ak_y) +\frac{2\sin^2(ak_y)\cos^2(ak_y)}{\left[\mathcal{Z}(k_x,k_y)\right]^2}\right]\right\}.
\end{align}

After utilizing $\frac{1}{x+\mathrm{i}\eta}=\frac{1}{x}-\mathrm{i}\pi\delta(x)$, the real part of the longitudinal optical conductivities (LOCs) are
\begin{align}
	&\mathrm{Re}~\sigma_{jj}(\omega,\mu,t,h)=\frac{\mathrm{i}}{\omega}\Pi_{jj}(\omega,\mu,t,h)
	=\frac{-\mathrm{Im}~\Pi_{jj}(\omega,\mu,t,h)}{\omega}\notag\\
	=&e^2\sum_{\lambda=\pm}\sum_{\lambda^{\prime}=\pm} \int_{-\frac{\pi}{2a}}^{+\frac{\pi}{2a}}\frac{\mathrm{d}k_x}{2\pi}\int_{-\frac{\pi}{a}}^{+\frac{\pi}{a}}\frac{\mathrm{d}k_y}{2\pi} \mathcal{F}_{jj}^{\lambda,\lambda^{\prime}}(\boldsymbol{k}) \frac{f\left[\varepsilon_{\lambda}(k_x,k_y)\right]-f\left[\varepsilon_{\lambda^{\prime}}(k_x,k_y)\right]} {\omega}\delta\left[\omega+(\lambda-\lambda^{\prime})\mathcal{Z}(k_x,k_y)\right].
\end{align}

We focus on the real part of the interband LOCs take the form
\begin{align}
	\mathrm{Re}~\sigma_{jj}^{\mathrm{IB}}(\omega,\mu,t,h)=& e^2\pi \int_{-\frac{\pi}{2a}}^{+\frac{\pi}{2a}}\frac{\mathrm{d}k_x}{2\pi}\int_{-\frac{\pi}{a}}^{+\frac{\pi}{a}}\frac{\mathrm{d}k_y}{2\pi} \mathcal{F}_{jj}^{-,+}(\boldsymbol{k}) \frac{f\left[\varepsilon_{-}(k_x,k_y)\right]-f\left[\varepsilon_{+}(k_x,k_y)\right]} {\omega}\delta\left[\omega-2\mathcal{Z}(k_x,k_y)\right].
\end{align}

\section{Analytical expressions of conventional critical frequencies and partner frequencies by utilizing the Lagrange multiplier method \label{Sec:AppendixB}}

The energy dispersion can be recast as
\begin{align}
	\tilde{\varepsilon}_{\lambda}^{\kappa}\left(\tilde{k}_{x},\tilde{k}_{y}\right)&=\kappa t \sin(a\tilde{k}_{y})\varepsilon_{0}-\kappa h\cos(a\tilde{k}_{y})\varepsilon_{0}+\lambda\sqrt{\sin^{2}(a\tilde{k}_{x})+\sin^{2}(a\tilde{k}_{y})}\varepsilon_{0}
	\nonumber\\
	&=\kappa t Y\varepsilon_{0}-\kappa h\sqrt{1-Y^{2}}\varepsilon_{0}+\lambda\sqrt{X^{2}+Y^{2}}\varepsilon_{0},
\end{align}
where 
\begin{align}
	\tilde{k}_{x}\equiv \tilde{k} \cos\tilde{\theta}_{k} &\in \left[-\frac{\pi}{2a},+\frac{\pi}{2a}\right],\\
	\tilde{k}_{y}\equiv \tilde{k} \sin\tilde{\theta}_{k}&\in \left[-\frac{\pi}{2a},+\frac{\pi}{2a}\right],
\end{align}
which lead to 
\begin{align}
	\sin(a\tilde{k}_{x})&=X\in \left[-1,+1\right],\\
	\cos(a\tilde{k}_{x})&=\sqrt{1-X^{2}}\in \left[0,+1\right],\\
	\sin(a\tilde{k}_{y})&=Y\in \left[-1,+1\right],\\
	\cos(a\tilde{k}_{y})&=\sqrt{1-Y^{2}}\in \left[0,+1\right].\\
\end{align}

%

For given $\lambda $ and $\kappa$, we find the extreme values by utilizing the Lagrange multiplier method
\begin{equation}
	\mathcal{L}=2\sqrt{X^{2}+Y^{2}}\varepsilon_{0}+\zeta\left(\kappa tY\varepsilon_{0}-\kappa h\sqrt{1-Y^{2}}\varepsilon_{0}+\lambda\sqrt{X^{2}+Y^{2}}\varepsilon_{0}-\mu\right),
\end{equation}
with $\zeta$ being the Lagrange multiplier. Equivalently, we have
\begin{align}
	\frac{\partial\mathcal{L}}{\partial X}= & \frac{\left(2+\lambda \zeta\right) X\varepsilon_{0}}{\sqrt{X^{2}+Y^{2}}}=0,
	\label{Condition1}
	\\
	\frac{\partial\mathcal{L}}{\partial Y}= & \zeta\kappa t\varepsilon_{0}+
	\zeta\kappa h\frac{Y}{\sqrt{1-Y^{2}}}\varepsilon_{0}
	+\left(2+\lambda \zeta\right)\frac{Y}{\sqrt{X^{2}+Y^{2}}}\varepsilon_{0}=0,
	\label{Condition2}
	\\
	\frac{\partial\mathcal{L}}{\partial \zeta}= & \kappa tY\varepsilon_{0}-\kappa h\sqrt{1-Y^{2}}\varepsilon_{0}+\lambda\sqrt{X^{2}+Y^{2}}\varepsilon_{0}-\mu=0.
	\label{Condition3}
\end{align}
It is noted that the condition in Eq.(\ref{Condition1}) is satisfied only when $X=0$ and $X\neq 0$, which will be detailedly discussed in the following two subsections.

\subsection{Analytical results for $X=0$}

For $X=0$, 
\begin{align}
	\sin\left(a\tilde{k}_{x}\right)=0,
\end{align}
or equivalently 
\begin{align}
	&\tilde{\theta}_{k}=\frac{\pi}{2},\\
	&\tilde{\theta}_{k}=\frac{3\pi}{2}.
\end{align}
We list the discussion as follows.

\subsubsection{Analytical results for $h=0$, $t=0$ and $0<\mu<(1-t)\varepsilon_{0}$}

If $h=0$, $t=0$ and $\mu>0$, the conditions in Eq.(\ref{Condition1}) and (\ref{Condition2}) are simultaneously satisfied only when $\left(2+\lambda\zeta\right)=0$. Then the condition in Eq.(\ref{Condition3}) gives rise to 
\begin{align}
	\sqrt{X^{2}+Y^{2}}\varepsilon_{0}=|Y|\varepsilon_{0}\equiv \frac{\omega}{2}&\ge 0,\\
	\lambda\frac{\omega}{2}-\mu&=0,
\end{align}
leading to 
\begin{align}
	\omega&=\frac{2\mu}{\lambda}\ge 0.
\end{align}

Since $\mu>0$, we get $\lambda=+1$, then 
\begin{align}
	\omega&\equiv 2\mu.
\end{align}


\subsubsection{Analytical results for $h=0$, $0<t<1$ and $0<\mu<(1-t)\varepsilon_{0}$}

From the solution $X=0$, the condition in Eq.(\ref{Condition1}) is automatically satisfied, and the condition in Eq.(\ref{Condition2}) can be satisfied if a suitable Lagrange multiplier $\zeta$ is chosen. Therefore, we focus on the condition in Eq.(\ref{Condition3}), which is independent of $\zeta$.
We have $\frac{\omega}{2}=\sqrt{X^2+Y^2}\varepsilon_{0}=|Y|\varepsilon_{0}=\mathrm{sgn}(Y)Y\varepsilon_{0}$ for $X=0$. It is straightforward that the condition in Eq.(\ref{Condition3}) reduces to 
\begin{align}
	\mathrm{sgn}(Y)\lambda \left[\kappa t Y+\mathrm{sgn}(Y)\lambda  Y \right]\varepsilon_{0}&=\mathrm{sgn}(Y)\lambda \mu
\end{align}

\begin{align}
	Y&=\frac{\mathrm{sgn}(Y)\lambda\frac{\mu}{\varepsilon_{0}}}{1+\mathrm{sgn}(Y)\kappa\lambda t}
\end{align}

\begin{align}
	\frac{\omega}{2}&=|Y|\varepsilon_{0}=\frac{\mu}{1+\mathrm{sgn}(Y)\kappa\lambda t}\ge 0
\end{align}

For $h=0$, $0<t<1$ and $\mu>0$, we have 
\begin{align}
	Y_{+}&=\frac{+\lambda\frac{\mu}{\varepsilon_{0}}}{1+\kappa\lambda t},\\
	Y_{-}&=\frac{-\lambda\frac{\mu}{\varepsilon_{0}}}{1-\kappa\lambda t}.
\end{align}

For $\kappa\lambda=+1$, we have
\begin{align}
	Y_{+}&=\frac{+\lambda\frac{\mu}{\varepsilon_{0}}}{1+ t};\\
	Y_{-}&=\frac{-\lambda\frac{\mu}{\varepsilon_{0}}}{1-t}.
\end{align}

For $\kappa\lambda=-1$, we have
\begin{align}
	Y_{+}&=\frac{+\lambda\frac{\mu}{\varepsilon_{0}}}{1- t};\\
	Y_{-}&=\frac{-\lambda\frac{\mu}{\varepsilon_{0}}}{1+t}.
\end{align}

When $h=0$, $0<t<1$ and $\mu>0$, the band index $\lambda$ must be $+1$. Consequently, we have: for $\kappa=+1$, 
\begin{align}
	Y_{+}&=\frac{+\frac{\mu}{\varepsilon_{0}}}{1+ t},\\
	Y_{-}&=\frac{-\frac{\mu}{\varepsilon_{0}}}{1-t},\\
	\omega_{2}^{+}
	&=2|Y_{-}|\varepsilon_{0}=\frac{2\mu}{1-t},\\
	\omega_{2}^{-}
	&=2|Y_{+}|\varepsilon_{0}=\frac{2\mu}{1+t},
\end{align}
and for $\kappa=-1$
\begin{align}
	Y_{+}&=\frac{+\frac{\mu}{\varepsilon_{0}}}{1-t},\\
	Y_{-}&=\frac{-\frac{\mu}{\varepsilon_{0}}}{1+t},\\
	\omega_{1}^{+}
	&=2|Y_{+}|\varepsilon_{0}=\frac{2\mu}{1-t},\\
	\omega_{1}^{-}
	&=2|Y_{-}|\varepsilon_{0}=\frac{2\mu}{1+t}.
\end{align}

\subsubsection{Analytical results for $h>0$ }


The condition in Eq.(\ref{Condition1}) is automatically satisfied, and the condition in Eq.(\ref{Condition2}) can be satisfied if a suitable Lagrange multiplier $\zeta$ is chosen. Therefore, we focus on the condition in Eq.(\ref{Condition3}), which is independent of $\zeta$.
We have $\sqrt{X^2+Y^2}=|Y|=\mathrm{sgn}(Y)Y$ for $X=0$. It is straightforward that the condition in Eq.(\ref{Condition3}) reduces to 
\begin{align}
	\kappa tY-\kappa h\sqrt{1-Y^{2}}+\mathrm{sgn}(Y)\lambda Y=\frac{\mu}{\varepsilon_{0}},
	\label{NewCon}
\end{align}

which leads to 
\begin{align}
	\left[\mathrm{sgn}(Y)\lambda+\kappa t\right]Y-\frac{\mu}{\varepsilon_{0}}=\kappa h\sqrt{1-Y^{2}},
\end{align}
namely,
\begin{align}
	\kappa\left[\mathrm{sgn}(Y)\lambda +\kappa t\right]Y-\kappa\frac{\mu}{\varepsilon_{0}}=
	\left[\mathrm{sgn}(Y)\kappa \lambda +t\right]Y-\kappa\frac{\mu}{\varepsilon_{0}}=h\sqrt{1-Y^{2}}\ge 0,
\end{align}

Consequently, we have
\begin{align}
	\left[\mathrm{sgn}(Y)\kappa \lambda +t\right]^2Y^2-2\kappa\frac{\mu}{\varepsilon_{0}}\left[\mathrm{sgn}(Y)\kappa \lambda +t\right]Y
	+\frac{\mu^2}{\varepsilon_{0}^2}&=h^2-h^2Y^{2},\\
	\left[\mathrm{sgn}(Y)\kappa \lambda +t\right]Y-\kappa\frac{\mu}{\varepsilon_{0}}&\ge 0.
\end{align}
which can further reduce to
\begin{align}
	\left\{\left[1+\mathrm{sgn}(Y)\kappa \lambda t\right]^2+h^2\right\}Y^2-2\mathrm{sgn}(Y)\lambda \frac{\mu}{\varepsilon_{0}}\left[1+\mathrm{sgn}(Y)\kappa \lambda t\right]Y
	-\left(h^2-\frac{\mu^2}{\varepsilon_{0}^2}\right)&=0,\\
	\mathrm{sgn}(Y)\kappa \lambda \left[1+\mathrm{sgn}(Y)\kappa \lambda t\right]Y&\ge\kappa~\frac{\mu}{\varepsilon_{0}}.
\end{align}

The solutions take
\begin{align}
	Y_{\pm}&=\frac{2\mathrm{sgn}(Y)\lambda \frac{\mu}{\varepsilon_{0}}\left[1+\mathrm{sgn}(Y)\kappa \lambda  t\right]\pm\sqrt{\left\{2\mathrm{sgn}(Y)\lambda \frac{\mu}{\varepsilon_{0}}\left[1+\mathrm{sgn}(Y)\kappa \lambda  t\right]\right\}^{2}+4\left\{\left[1+\mathrm{sgn}(Y)\kappa \lambda  t\right]^2+h^2\right\}\left(h^2-\frac{\mu^2}{\varepsilon_{0}^2}
			\right)}}{2\left\{\left[1+\mathrm{sgn}(Y)\kappa \lambda  t\right]^2+h^2\right\}}\nonumber\\
	&=\pm\frac{\sqrt{4h^2\left\{\left[1+\mathrm{sgn}(Y)\kappa \lambda  t\right]^{2}+\left(h^2-\frac{\mu^2}{\varepsilon_{0}^2}\right)\right\}}\pm 2\mathrm{sgn}(Y)\lambda \frac{\mu}{\varepsilon_{0}}\left[1+\mathrm{sgn}(Y)\kappa \lambda  t\right]}{2\left\{\left[1+\mathrm{sgn}(Y)\kappa \lambda  t\right]^2+h^2\right\}}\nonumber\\
	&=\pm\frac{4h^2\left\{\left[1+\mathrm{sgn}(Y)\kappa \lambda  t\right]^{2}+h^2-\frac{\mu^2}{\varepsilon_{0}^2}\right\}- 4\frac{\mu^2}{\varepsilon_{0}^2}\left[1+\mathrm{sgn}(Y)\kappa \lambda  t\right]^2}{2\left\{\left[1+\mathrm{sgn}(Y)\kappa \lambda  t\right]^2+h^2\right\}\left\{\sqrt{4h^2\left\{\left[1+\mathrm{sgn}(Y)\kappa \lambda  t\right]^{2}+\left(h^2-\frac{\mu^2}{\varepsilon_{0}^2}\right)\right\}}\mp 2\mathrm{sgn}(Y)\lambda \frac{\mu}{\varepsilon_{0}}\left[1+\mathrm{sgn}(Y)\kappa \lambda  t\right]\right\}}\nonumber\\
	&=\pm\frac{4\left\{\left[1+\mathrm{sgn}(Y)\kappa \lambda  t\right]^{2}+h^2\right\}\left(h^2-\frac{\mu^2}{\varepsilon_{0}^2}\right)}{2\left\{\left[1+\mathrm{sgn}(Y)\kappa \lambda  t\right]^2+h^2\right\}\left\{\sqrt{4h^2\left\{\left[1+\mathrm{sgn}(Y)\kappa \lambda  t\right]^{2}+\left(h^2-\frac{\mu^2}{\varepsilon_{0}^2}\right)\right\}}\mp 2\mathrm{sgn}(Y)\lambda \frac{\mu}{\varepsilon_{0}}\left[1+\mathrm{sgn}(Y)\kappa \lambda  t\right]\right\}}\nonumber\\
	&=\pm\frac{\left(h^2-\frac{\mu^2}{\varepsilon_{0}^2}\right)}{\sqrt{h^2\left\{\left[1+\mathrm{sgn}(Y)\kappa \lambda  t\right]^{2}+\left(h^2-\frac{\mu^2}{\varepsilon_{0}^2}\right)\right\}}\mp \mathrm{sgn}(Y)\lambda \frac{\mu}{\varepsilon_{0}}\left[1+\mathrm{sgn}(Y)\kappa \lambda  t\right]},
\end{align}
and
\begin{align}
	&\mathrm{sgn}(Y_{\pm})\kappa\lambda\left[1+\mathrm{sgn}(Y_{\pm})\kappa\lambda t\right]Y_{\pm}-\kappa\frac{\mu}{\varepsilon_{0}}\ge 0.
\end{align}

\subsubsection{Analytical results for $h>0$, $0\le t<1$ and $\mu=0$}

For $h>0$, $0\le t<1$ and $\mu=0$, we have 
\begin{align}
	Y_{+}&=+\frac{h}{\sqrt{\left(1+\kappa\lambda t\right)^2+h^2}},\hspace{1cm}\frac{+\kappa\lambda\left(1+\kappa\lambda t\right)h}{\sqrt{\left(1+\kappa\lambda t\right)^2+h^2}}\ge 0;\\
	Y_{-}&=-\frac{h}{\sqrt{\left(1-\kappa\lambda t\right)^2+h^2}},
	\hspace{1cm}\frac{-\kappa\lambda\left(1-\kappa\lambda t\right)h}{\sqrt{\left(1-\kappa\lambda t\right)^2+h^2}}\ge0.
\end{align}

For $\kappa\lambda=+1$, we have
\begin{align}
	Y_{+}&=+\frac{h}{\sqrt{\left(1+ t\right)^2+h^2}},\hspace{1cm}\frac{\left(1+ t\right)h}{\sqrt{\left(1+t\right)^2+h^2}}\ge 0;\\
	Y_{-}&=-\frac{h}{\sqrt{\left(1-t\right)^2+h^2}},
	\hspace{1cm}\frac{-\left(1- t\right)h}{\sqrt{\left(1-t\right)^2+h^2}}< 0~(\ngeq0).
\end{align}
If $(\kappa,\lambda)=(+1,+1)$ or $(\kappa,\lambda)=(-1,-1)$, we further have
\begin{align}
	Y_{+}&=+\frac{h}{\sqrt{\left(1+ t\right)^2+h^2}},\\
	\omega_{1}^{\pm}&=2|Y_{+}|\varepsilon_{0}=+\frac{2h\varepsilon_{0}}{\sqrt{\left(1+ t\right)^2+h^2}},\\
	\omega_{2}^{\pm}&=2|Y_{+}|\varepsilon_{0}=+\frac{2h\varepsilon_{0}}{\sqrt{\left(1+ t\right)^2+h^2}}.
\end{align}

\subsubsection{Analytical results for $h>0$, $0\le t<1$ and $0< \mu\le\frac{h}{\sqrt{1+h^2}}\varepsilon_{0}<\frac{h}{\sqrt{t^2+h^2}}\varepsilon_{0}$}

From the expressions
\begin{align}
	Y_{\pm}&=\pm\frac{\left(h^2-\frac{\mu^2}{\varepsilon_{0}^2}\right)}{\sqrt{h^2\left\{\left[1+\mathrm{sgn}(Y_{\pm})\kappa\lambda t\right]^{2}+\left(h^2-\frac{\mu^2}{\varepsilon_{0}^2}\right)\right\}}\mp \mathrm{sgn}(Y_{\pm})\lambda\frac{\mu}{\varepsilon_{0}}\left[1+\mathrm{sgn}(Y_{\pm})\kappa\lambda t\right]},
\end{align}
and
\begin{align}
	&\mathrm{sgn}(Y_{\pm})\kappa\lambda\left[1+\mathrm{sgn}(Y_{\pm})\kappa\lambda t\right]Y_{\pm}-\kappa\frac{\mu}{\varepsilon_{0}}\ge 0,
\end{align}
we obtain that, for $h>0$, $0\le t<1$ and $0\le \mu\le\frac{h}{\sqrt{1+h^2}}\varepsilon_{0}<\frac{h}{\sqrt{t^2+h^2}}\varepsilon_{0}$,
\begin{align}
	\left(h^2-\frac{\mu^2}{\varepsilon_{0}^2}\right)&>0,\\
	\sqrt{4h^2\left\{\left[1+\mathrm{sgn}(Y_{\pm})\kappa\lambda  t\right]^{2}+\left(h^2-\frac{\mu^2}{\varepsilon_{0}^2}\right)\right\}}&> \left|2\mathrm{sgn}(Y_{\pm})\lambda \frac{\mu}{\varepsilon_{0}}\left[1+\mathrm{sgn}(Y_{\pm})\kappa\lambda  t\right]\right|
	=2\frac{\mu}{\varepsilon_{0}}\left|1+\mathrm{sgn}(Y_{\pm})\kappa\lambda  t\right|.
\end{align}
Consequently, we have 
\begin{align}
	\mathrm{sgn}\left(Y_{\pm}\right)&=\pm,
\end{align}
and
\begin{align}
	&\pm\kappa\lambda\left(1\pm\kappa\lambda t\right)Y_{\pm}-\kappa\frac{\mu}{\varepsilon_{0}}\ge 0.
\end{align}

Therefore,
\begin{align}
	Y_{\pm}&=\pm\frac{\left(h^2-\frac{\mu^2}{\varepsilon_{0}^2}\right)}{\sqrt{h^2\left\{\left(1\pm\kappa\lambda t\right)^{2}+\left(h^2-\frac{\mu^2}{\varepsilon_{0}^2}\right)\right\}}- \lambda\frac{\mu}{\varepsilon_{0}}\left(1\pm\kappa\lambda t\right)},
\end{align}
and
\begin{align}
	&\pm\kappa\lambda\left(1\pm\kappa\lambda t\right)Y_{\pm}-\kappa\frac{\mu}{\varepsilon_{0}}\ge 0.
\end{align}

\begin{align}
	Y_{\pm}&=\pm\frac{\left(h^2-\frac{\mu^2}{\varepsilon_{0}^2}\right)}{\sqrt{h^2\left\{\left(1\pm\kappa\lambda t\right)^{2}+\left(h^2-\frac{\mu^2}{\varepsilon_{0}^2}\right)\right\}}- \lambda\frac{\mu}{\varepsilon_{0}}\left(1\pm\kappa\lambda t\right)}\nonumber\\
	&=\pm\frac{\sqrt{\frac{\left(h^2-\frac{\mu^2}{\varepsilon_{0}^2}\right)}{h^2}}}{\sqrt{1+\frac{\left(1\pm\kappa\lambda t\right)^{2}}{\left(h^2-\frac{\mu^2}{\varepsilon_{0}^2}\right)}}- \lambda\frac{\mu}{\varepsilon_{0}h}\sqrt{\frac{\left(1\pm\kappa\lambda t\right)^2}{\left(h^2-\frac{\mu^2}{\varepsilon_{0}^2}\right)}}},
\end{align}
and
\begin{align}
	&\pm\kappa\lambda\left(1\pm\kappa\lambda t\right)Y_{\pm}-\kappa\frac{\mu}{\varepsilon_{0}}\ge 0.
\end{align}

For $\kappa\lambda=+1$, we have
\begin{align}
	Y_{\pm}&=\pm\frac{\left(h^2-\frac{\mu^2}{\varepsilon_{0}^2}\right)}{\sqrt{h^2\left\{\left(1\pm t\right)^{2}+\left(h^2-\frac{\mu^2}{\varepsilon_{0}^2}\right)\right\}}- \lambda\frac{\mu}{\varepsilon_{0}}\left(1\pm t\right)},
\end{align}
and
\begin{align}
	&\left(1\pm t\right)\frac{\left(h^2-\frac{\mu^2}{\varepsilon_{0}^2}\right)}{\sqrt{h^2\left\{\left(1\pm t\right)^{2}+\left(h^2-\frac{\mu^2}{\varepsilon_{0}^2}\right)\right\}}- \lambda\frac{\mu}{\varepsilon_{0}}\left(1\pm t\right)}-\kappa\frac{\mu}{\varepsilon_{0}}\ge 0.
\end{align}

If $(\kappa,\lambda)=(+1,+1)$, we have
\begin{align}
	Y_{\pm}&=\pm\frac{\left(h^2-\frac{\mu^2}{\varepsilon_{0}^2}\right)}{\sqrt{h^2\left\{\left(1\pm t\right)^{2}+\left(h^2-\frac{\mu^2}{\varepsilon_{0}^2}\right)\right\}}- \frac{\mu}{\varepsilon_{0}}\left(1\pm t\right)},
\end{align}
and
\begin{align}
	&\left(1\pm t\right)\frac{\left(h^2-\frac{\mu^2}{\varepsilon_{0}^2}\right)}{\sqrt{h^2\left\{\left(1\pm t\right)^{2}+\left(h^2-\frac{\mu^2}{\varepsilon_{0}^2}\right)\right\}}- \frac{\mu}{\varepsilon_{0}}\left(1\pm t\right)}-\frac{\mu}{\varepsilon_{0}}\ge 0.
\end{align}

\begin{align}
	\omega_{2}^{+}&=2|Y_{-}|\varepsilon_{0}=\frac{2\left(h^2-\frac{\mu^2}{\varepsilon_{0}^2}\right)}{\sqrt{h^2\left\{\left(1- t\right)^{2}+\left(h^2-\frac{\mu^2}{\varepsilon_{0}^2}\right)\right\}}- \frac{\mu}{\varepsilon_{0}}\left(1-t\right)}\varepsilon_{0},\\
	\omega_{2}^{-}&=2|Y_{+}|\varepsilon_{0}=\frac{2\left(h^2-\frac{\mu^2}{\varepsilon_{0}^2}\right)}{\sqrt{h^2\left\{\left(1+ t\right)^{2}+\left(h^2-\frac{\mu^2}{\varepsilon_{0}^2}\right)\right\}}- \frac{\mu}{\varepsilon_{0}}\left(1+t\right)}\varepsilon_{0}.
\end{align}

If $(\kappa,\lambda)=(-1,-1)$, we have
\begin{align}
	Y_{\pm}&=\pm\frac{\left(h^2-\frac{\mu^2}{\varepsilon_{0}^2}\right)}{\sqrt{h^2\left\{\left(1\pm t\right)^{2}+\left(h^2-\frac{\mu^2}{\varepsilon_{0}^2}\right)\right\}}+\frac{\mu}{\varepsilon_{0}}\left(1\pm t\right)},
\end{align}
and
\begin{align}
	&\left(1\pm t\right)\frac{\left(h^2-\frac{\mu^2}{\varepsilon_{0}^2}\right)}{\sqrt{h^2\left\{\left(1\pm t\right)^{2}+\left(h^2-\frac{\mu^2}{\varepsilon_{0}^2}\right)\right\}}+\frac{\mu}{\varepsilon_{0}}\left(1\pm t\right)}+\frac{\mu}{\varepsilon_{0}}\ge 0.
\end{align}
\begin{align}
	\omega_{1}^{+}&=2|Y_{-}|\varepsilon_{0}=\frac{2\left(h^2-\frac{\mu^2}{\varepsilon_{0}^2}\right)}{\sqrt{h^2\left\{\left(1- t\right)^{2}+\left(h^2-\frac{\mu^2}{\varepsilon_{0}^2}\right)\right\}}+\frac{\mu}{\varepsilon_{0}}\left(1-t\right)}\varepsilon_{0},\\
	\omega_{1}^{-}&=2|Y_{+}|\varepsilon_{0}=\frac{2\left(h^2-\frac{\mu^2}{\varepsilon_{0}^2}\right)}{\sqrt{h^2\left\{\left(1+ t\right)^{2}+\left(h^2-\frac{\mu^2}{\varepsilon_{0}^2}\right)\right\}}+\frac{\mu}{\varepsilon_{0}}\left(1+t\right)}\varepsilon_{0}.
\end{align}

\subsection{Analytical results for $X\neq 0$}

For $X\neq 0$, the condition in Eq.(\ref{Condition1}) requires that $\left(2+\lambda \zeta\right)=0$. Consequently, the condition in Eq.(\ref{Condition2}) reduces to 
\begin{align}
	t+h\frac{Y}{\sqrt{1-Y^{2}}}=0.
	\label{partnerfre0}
\end{align}

\subsubsection{Analytical results for $h=0$, $t=0$ and $\mu>0$}

For $h=0$, $t=0$ and $\mu>0$, the condition in Eq.(\ref{partnerfre0}) is automatically satisfied, and $Y$ is not constrained. 

\begin{align}
	\omega^{\prime}&=2\sqrt{X^2+Y^2}\varepsilon_{0}=2\lambda\mu
	\ge 0.
\end{align}
For $\mu>0$, we have $\lambda=+1$. As a consequence, 
\begin{align}
	\omega_{j}^{\prime}&=2\mu.
\end{align}

\subsubsection{Analytical results for $h>0$, $t=0$ and $\mu\ge 0$}

For $h>0$, $t=0$ and $\mu\ge 0$, the condition in Eq.(\ref{partnerfre0}) requires $Y=0$, and hence we have the partner frequencies 
\begin{align}
	\omega^{\prime}=2\sqrt{X^2+Y^2}\varepsilon_{0}&=2|Y|\varepsilon_{0}=2\kappa\lambda\left[h\varepsilon_{0}+\kappa\mu\right]\ge 0.
\end{align}

For $0<\mu<h\varepsilon_{0}$, if $\kappa\lambda=+1$, 
\begin{align}
	\omega^{\prime}&=2\left[h\varepsilon_{0}+\kappa\mu\right]\ge 0,
\end{align}
if $\kappa\lambda=-1$, 
\begin{align}
	\omega^{\prime}&=-2\left[h\varepsilon_{0}+\kappa\mu\right]<0~(\ngeq 0).
\end{align}

In brief, for $t=0$, $h>0$ and $0<\mu<\sqrt{t^2+h^2}\varepsilon_{0}$, if $\kappa\lambda=+1$, 
\begin{align}
	\omega^{\prime}=2\sqrt{X^2+Y^2}\varepsilon_{0}&=2\kappa\lambda\left[\sqrt{h^2+t^{2}}\varepsilon_{0}+\kappa\mu\right]\ge 0.
\end{align}

For $(\kappa,\lambda)=(-1,-1)$
\begin{align}
	\omega_{1}^{\prime}&=2\left|h\varepsilon_{0}-\mu\right|,\hspace{1cm}h\varepsilon_{0}-\mu\ge 0,
\end{align}
and for $(\kappa,\lambda)=(+1,+1)$,
\begin{align}
	\omega_{2}^{\prime}&=2\left|h\varepsilon_{0}+\mu\right| ,\hspace{1cm}h\varepsilon_{0}+\mu\ge 0.
\end{align}

In brief, 
\begin{align}
	\omega_{j}^{\prime}&=2\left|h\varepsilon_{0}+(-1)^{j}\mu\right| ,\hspace{1cm}h\varepsilon_{0}+(-1)^{j}\mu\ge 0.
\end{align}

\subsubsection{Analytical results for $h=0$, $0<t<1$ and $\mu>0$}

For $h=0$, $0<t<1$ and $\mu>0$, the conditions in Eq.(\ref{Condition1}) and Eq.(\ref{Condition2})---or equivalently the condition in Eq.(\ref{partnerfre0})---can not be satisfied, indicating that there is no extreme value of $\omega_{j}^{\prime}$.

\subsubsection{Analytical results for $h>0$, $0<t<1$ and $\mu\ge 0$}

For $h>0$ and $0<t<1$, if $-1<Y<1$, we have
\begin{align}
	\sqrt{1-Y^{2}}=-\frac{h}{t}Y\ge 0.
\end{align}
Further, we have
\begin{align}
	&-1<Y\le 0,\\
	&1-Y^{2}=\frac{h^2}{t^2}Y^2.
\end{align}

The solution takes
\begin{align}
	&Y=-\frac{t}{\sqrt{h^2+t^2}}.
\end{align}

From the condition in Eq.(\ref{Condition3}), 
\begin{align}
	&\sqrt{X^2+Y^2}\varepsilon_{0}=\lambda\left[\mu-\kappa tY\varepsilon_{0}+\kappa h\sqrt{1-Y^{2}}\varepsilon_{0}\right]
	=\lambda\left[\mu+\kappa \sqrt{h^2+t^{2}}\varepsilon_{0}\right]
	=\kappa\lambda\left[\sqrt{h^2+t^{2}}\varepsilon_{0}+\kappa\mu\right]
	\ge 0.
\end{align}

Finally, we have the partner frequencies 
\begin{align}
	\omega^{\prime}=2\sqrt{X^2+Y^2}\varepsilon_{0}&=2\kappa\lambda\left[\sqrt{h^2+t^{2}}\varepsilon_{0}+\kappa\mu\right]\ge 0.
\end{align}

For $0<\mu<\sqrt{t^2+h^2}\varepsilon_{0}$, if $\kappa\lambda=+1$, 
\begin{align}
	\omega^{\prime}=2\sqrt{X^2+Y^2}\varepsilon_{0}&=2\left[\sqrt{h^2+t^{2}}\varepsilon_{0}+\kappa\mu\right]\ge 0,
\end{align}
if $\kappa\lambda=-1$, 
\begin{align}
	\omega^{\prime}=2\sqrt{X^2+Y^2}\varepsilon_{0}&=-2\left[\sqrt{h^2+t^{2}}\varepsilon_{0}+\kappa\mu\right]<0~(\ngeq 0).
\end{align}

In brief, for $0\le t<1$, $h>0$ and $0<\mu<\sqrt{t^2+h^2}\varepsilon_{0}$, if $\kappa\lambda=+1$, 
\begin{align}
	\omega^{\prime}=2\sqrt{X^2+Y^2}\varepsilon_{0}&=2\kappa\lambda\left[\sqrt{h^2+t^{2}}\varepsilon_{0}+\kappa\mu\right]\ge 0.
\end{align}

For $(\kappa,\lambda)=(+1,+1)$,
\begin{align}
	\omega_{2}^{\prime}&=2\left|\sqrt{h^2+t^{2}}\varepsilon_{0}+\mu\right| ,\hspace{1cm}\sqrt{h^2+t^{2}}\varepsilon_{0}+\mu\ge 0,
\end{align}
and for $(\kappa,\lambda)=(-1,-1)$
\begin{align}
	\omega_{1}^{\prime}&=2\left|\sqrt{h^2+t^{2}}\varepsilon_{0}-\mu\right|,\hspace{1cm}\sqrt{h^2+t^{2}}\varepsilon_{0}-\mu\ge 0.
\end{align}

\subsection{Explicit values of conventional critical frequencies and partner frequencies}

In the following, we provide explicit values of both the conventional critical frequencies and their partner frequencies, as specified by the parameters in the main text. This is done to demonstrate quantitative consistency between the analytical expressions derived in the preceding three subsections and the full numerical calculations in the main text.

\subsubsection{untilted cases}

When $t=0.0,h=0.0,\mu=0.45~\varepsilon_{0}$,
\begin{align}
	\omega_{1}^{\pm}&=\frac{2\mu}{1\mp t}\equiv 
	\omega_{1}=0.9~\varepsilon_{0},\\
	\omega_{2}^{\pm}&=\frac{2\mu}{1\mp t}\equiv \omega_{2}=0.9~\varepsilon_{0},\\
	\omega_{1}^{\prime}&=2\mu=0.9~\varepsilon_{0},\\
	\omega_{2}^{\prime}&=2\mu=0.9~\varepsilon_{0}.
\end{align}

When $t=0.0,h=0.6,\mu=0.0~\varepsilon_{0}$,
\begin{align}
	\omega_{1}^{\pm}&=\frac{2h\varepsilon_{0}}{\sqrt{\left(1+t\right)+h^2}}\equiv 
	\omega_{1}=1.02899~\varepsilon_{0},\\
	\omega_{2}^{\pm}&=\frac{2h\varepsilon_{0}}{\sqrt{\left(1+t\right)+h^2}}\equiv \omega_{2}=1.02899~\varepsilon_{0},\\
	\omega_{1}^{\prime}&=2h\varepsilon_{0}=1.2~\varepsilon_{0},\\
	\omega_{2}^{\prime}&=2h\varepsilon_{0}=1.2~\varepsilon_{0}.
\end{align}

When $t=0.0,h=0.6,\mu=0.15~\varepsilon_{0}$, we have

\begin{align}
	\omega_{1}^{+}&=1.24103~\varepsilon_{0},\\
	\omega_{1}^{-}&=1.24103~\varepsilon_{0},\\
	\omega_{2}^{+}&=0.799856~\varepsilon_{0},\\
	\omega_{2}^{-}&=0.799856~\varepsilon_{0},\\
	\omega_{1}^{\prime}&=2\left[h\varepsilon_{0}+\mu\right]=1.5~\varepsilon_{0},\\
	\omega_{2}^{\prime}&=2\left[h\varepsilon_{0}-\mu\right]=0.9~\varepsilon_{0}.
\end{align}


\subsubsection{tilted cases}

When $t=0.3,h=0.0,\mu=0.45~\varepsilon_{0}$, we have

\begin{align}
	\omega_{1}^{+}&=1.28571~\varepsilon_{0},\\
	\omega_{1}^{-}&=0.69231~\varepsilon_{0},\\
	\omega_{2}^{+}&=1.28571~\varepsilon_{0},\\
	\omega_{2}^{-}&=0.69231~\varepsilon_{0}.
\end{align}

It is emphasized that there is no extreme value of $\omega_{j}^{\prime}$ in this case.

When $t=0.3,h=0.6,\mu=0.0~\varepsilon_{0}$, we have
\begin{align}
	\omega_{1}^{+}&=1.30158~\varepsilon_{0},\\
	\omega_{1}^{-}&=0.838116~\varepsilon_{0},\\
	\omega_{2}^{+}&=1.30158~\varepsilon_{0},\\
	\omega_{2}^{-}&=0.838116~\varepsilon_{0},\\
	\omega_{1}^{\prime}&=2\sqrt{h^2+t^{2}}\varepsilon_{0}=1.34164~\varepsilon_{0},\\
	\omega_{2}^{\prime}&=2\sqrt{h^2+t^{2}}\varepsilon_{0}=1.34164~\varepsilon_{0}.
\end{align}

When $t=0.3,h=0.6,\mu=0.15~\varepsilon_{0}$, we have
\begin{align}
	\omega_{1}^{+}&=1.53130~\varepsilon_{0},\\
	\omega_{1}^{-}&=1.02375~\varepsilon_{0},\\
	\omega_{2}^{+}&=1.03718~\varepsilon_{0},\\
	\omega_{2}^{-}&=0.64326~\varepsilon_{0},\\
	\omega_{1}^{\prime}&=2\left[\sqrt{h^2+t^{2}}\varepsilon_{0}+\mu\right]=1.64164~\varepsilon_{0},\\
	\omega_{2}^{\prime}&=2\left[\sqrt{h^2+t^{2}}\varepsilon_{0}-\mu\right]=1.04164~\varepsilon_{0}.
\end{align}
	
\end{widetext}

\end{document}